\documentclass[a4paper,fleqn,usenatbib]{mnras}

\usepackage{newtxtext,newtxmath}

\usepackage[T1]{fontenc}
\usepackage{ae,aecompl}

\usepackage{graphicx}	
\usepackage{amsmath}	
\usepackage{amssymb}	

\usepackage{textcomp}
\usepackage{lscape}
\usepackage{hyperref}

\usepackage{afterpage}

\newcommand{\chisq}{\chi^2}
\newcommand{\Oii}{\textsc{O~ii}}
\newcommand{\Oiii}{\textsc{O~iii}}
\newcommand{\Mratio}{\log(\rm M_{*}/M_{\odot})}
\newcommand{\Mdyn}{\rm M_{\rm dyn}}
\newcommand{\Mast}{\rm M_{\ast}}
\newcommand{\Msun}{\rm M_{\odot}}
\newcommand{\kms}{\rm km~s^{-1}}

\title[Stellar and Dynamical Masses of Cluster LCBGs at z$\sim$0.54]{Star-forming Galaxies in Intermediate Redshift Clusters: Stellar vs. Dynamical Masses of Luminous Compact Blue Galaxies}

\author[S. M. Randriamampandry et al.]
{S. M. Randriamampandry,$^{1,2,3}$\thanks{E-mail: solohery@saao.ac.za} S.~M. Crawford,$^{2,4}$  M.~A. Bershady,$^{6}$ G.~D. Wirth,$^{7}$ \newauthor and  C.~M. Cress$^{3,5}$ \\
$^{1}$Astrophysics \& Cosmology Research Unit, School of Maths, Statistics \& Computer Science, University of KwaZulu-Natal, \\ Durban 4041, South Africa \\ 
$^{2}$South African Astronomical Observatory, P.O. Box 9, Observatory 7935, Cape Town, South Africa\\
$^{3}$Department of Physics, University of the Western Cape, Private Bag X17, Bellville 7535, Cape Town, South Africa\\
$^{4}$Southern African Large Telescope, P.O. Box 9, Observatory 7935, Cape Town, South Africa\\
$^{5}$Centre For High Performance Computing, CSIR Campus, 15 Lower Hope St., Rosebank, Cape Town, South Africa \\
$^{6}$University of Wisconsin-Madison, 475 North Charter Street Madison, WI 53706, USA\\
$^{7}$W.~M. Keck Observatory, 65-1120 Mamalahoa Hwy, Kamuela, HI 96743, USA}

\date{Accepted XXX. Received YYY; in original form ZZZ}

\pubyear{2017}

\begin{document}
\label{firstpage}
\pagerange{\pageref{firstpage}--\pageref{lastpage}}
\maketitle

\begin{abstract}

We investigate the stellar masses of the class of star-forming objects known as Luminous Compact Blue Galaxies (LCBGs) by studying a sample of galaxies in the distant cluster MS\,0451.6-0305 at $z\approx0.54$ with ground-based multicolor imaging and spectroscopy.  For a sample of 16 spectroscopically-confirmed cluster LCBGs (colour $B-V < 0.5$, surface brightness $\mu_B < 21$ mag arcsec$^{-2}$, and magnitude $M_B < -18.5$), we measure stellar masses by fitting spectral energy distribution (SED) models to multiband photometry, and compare with dynamical masses (determined from velocity dispersion between 10 $<$ $\sigma_v  ~(\kms)$ $<$ 80), we previously obtained from their emission-line spectra.  We compare two different stellar population models that measure stellar mass in star-bursting galaxies, indicating correlations between the stellar age, extinction, and stellar mass derived from the two different SED models.  The stellar masses of cluster LCBGs are distributed similarly to those of field LCBGs, but the cluster LCBGs show lower dynamical-to-stellar mass ratios ($\Mdyn/\Mast = 2.6$) than their field LCBG counterparts ($\Mdyn/\Mast=4.8$), echoing trends noted previously in low-redshift dwarf elliptical galaxies. Within this limited sample, the specific star formation rate declines steeply with increasing mass, suggesting that these cluster LCBGs have undergone vigorous star formation.

\end{abstract}

\begin{keywords}
galaxies: -- cluster -- star-forming -- SEDs model  -- masses: stellar -- dynamical -- dark matter
\end{keywords}



\section{Introduction}  
 
Mass is an important tracer of the evolution of galaxies. Stellar masses provide an estimate of the integrated star formation history of a galaxy \citep{2000ApJ...536L..77B,2004Natur.428..625H,2005ApJ...622L...5T}, and can be used to trace how galaxies evolve over time.  For vigorously star-forming galaxies --- and, in particular, the special subclass of these objects known as Luminous Compact Blue Galaxies  \cite[LCBGs; see, e.g.,][]{1994ApJ...427L...9K,1997ApJ...478L..49K,2006ApJ...636L..13C} --- studying the relationship between stellar mass and dynamical mass at intermediate redshifts may illuminate possible evolutionary paths leading to the low-mass galaxies observed today.

LCBGs are characterized by their compactness, extreme blue colors, high rates of ongoing star formation, and diversity of morphologies \citep{1994ApJ...427L...9K,1997ApJ...489..543P, 2001ApJ...550..570H, 2004ApJ...615..689G,2006ApJ...640L.143N, 2007ApJ...671..310G, 2007AJ....134.2455H, 2010ApJ...708.1076T}. Previous work has shown that the LCBG population evolves rapidly over time,  their number density having dropped precipitously since $z\sim1$ in sync with the decrease in global star formation rate \citep{1997ApJ...489..559G}. In a series of papers \citep{2006ApJ...636L..13C,2011ApJ...741...98C,2016ApJ...817...87C}, we have identified and described the properties of LCBGs found in distant clusters.  Our key finding is that their properties (metalicity, dynamical mass, and size) and number density make LCBGs likely progenitors of the dwarf elliptical (dE) galaxies that dominate today's clusters.  By comparing the properties of field and cluster LCBGs such as masses and sizes, we can establish if these are the late-type progenitor galaxies.

Dwarf elliptical galaxies are a heterogeneous class of galaxies \citep{2009AN....330.1043L, 2012ApJS..198....2K} that are the most numerous type of galaxies in clusters although they are almost completely absent in the field. 
Previous studies have shown that some dwarf cluster galaxies have experienced a brief burst of star formation prior to being quenched \citep{2008MNRAS.385.1374M,  2014A&A...563A..27L, 2015MNRAS.452.1888R, 2016MNRAS.463.2819M}, although they do have a range of star formation histories \citep{2011MNRAS.417.1643K, 2012MNRAS.419.3167S, 2014ApJS..215...17T}. In addition, dwarf ellipticals show a range of complex photometric \citep{2012ApJ...745L..24J, 2014ApJ...786..105J} and kinematic structures \citep{2002MNRAS.332L..59P, 2012A&A...548A..78T, 2013MNRAS.428.2980R, 2014MNRAS.439..284R, 2014ApJ...783..120T, 2015ApJ...799..172T}.   Many of these studies point to the transformation of in-falling late-type galaxies by the cluster environment as being the progenitors of dwarf elliptical galaxies.  

Galaxy masses can be derived in two fundamentally different ways: Stellar masses are calculated from the spectral energy distribution of a galaxy's light by estimating the total number of stars contributing to its luminosity. Dynamical masses are determined from the characteristic size and internal velocity to estimate the total mass of the system. While stellar and dynamical masses are not necessarily equivalent, they use independent observational information and must satisfy the condition that the stellar mass is less than or equal to the total dynamical mass. Hence, these two measures of mass provide consistency checks on measurement systematics; with constraints on these systematics, differences between the two measures can reveal fundamental attributes such as the relative baryonic-to-dark-matter content of galaxies.  We can also exploit this attribute to test the hypothesis that LCBGs are potential progenitors of today's dE population. To facilitate this test, one of the aims of this work is to understand the limits on systematic error in stellar mass measurements for LCBGs.

A previous study has derived stellar and dynamical masses for LCBGs in the field \citep{2003ApJ...586L..45G}. We adopt this study to define LCBG properties in the field (hereafter {\it field sample}). In this work, we apply the same method to measure stellar and dynamical masses of LCBGs in galaxy clusters for the first time. This enables us to make a robust comparison of the effect of environment on stellar and dynamical masses of this class of galaxies at the same epoch.  For our cluster sample, we selected LCBGs from the massive, intermediate-redshift cluster MS\,0451.6-0305 (hereafter MS\,0451-03).  Observed at redshift $z\approx 0.538$, this cluster features a high velocity dispersion $\sigma = 1,354~\kms$, a large radius $\rm R_{200} = 2.5~\rm Mpc$, and X-ray luminosity $\rm L_{\rm x}^{\rm bol} = 4 \times 10^{45}~\rm erg~s^{-1}$ \cite[see][]{2003ApJ...598..190D, 2009ApJ...690.1158C}.

This paper is organized as follows. Section~\ref{sec:Observations and Data Analysis} describes our observations and data analysis as well as the galaxy sample selection. Section~\ref{sec:Stellar Mass Estimates} presents our stellar mass measurements based on SED models computed using two public stellar population synthesis (SPS) codes. Section~\ref{sec: Results} reports our measurements of stellar and dynamical masses,  specific star formation rates (sSFR), and stellar ages of the LCBG population. In Sections~\ref{sec:Discussion} and \ref{sec:Conclusion}, we discuss our results and then summarize our findings. Throughout this work, we adopt $\rm H_{0} = 71~\kms~{\rm Mpc}^{-1}$, $\Omega_{\rm m}$ = 0.27, and $\Omega_{\rm DE}$ = 0.73; all photometry is in the Vega magnitude system.

\begin{table*}
\centering
\caption{Optical and near-IR photometry for star-forming galaxies with secure spectroscopic redshifts.} 
\label{tab:photometry} 
{\scriptsize
\begin{tabular}{rccccccccc}\hline \hline 
{\footnotesize Obj} & {\footnotesize WLTV} & {\footnotesize $\alpha_{\rm J2000}$} & {\footnotesize $\delta_{\rm J2000}$} & {\footnotesize $U$} & {\footnotesize $B$} & {\footnotesize $R$} & {\footnotesize $I$} & {\footnotesize $z$} & {\footnotesize $K$}  \\ 
{\footnotesize ID} & {\footnotesize ID} & {\footnotesize (deg)} & (deg) & {\footnotesize (mag)} & {\footnotesize (mag)} & {\footnotesize (mag)} & {\footnotesize (mag)} & {\footnotesize (mag)} & {\footnotesize (mag)} \\
{\footnotesize (1)} & {\footnotesize (2)} & {\footnotesize (3)} & {\footnotesize (4)} & {\footnotesize (5)} & {\footnotesize (6)} & {\footnotesize (7)} & {\footnotesize (8)} & {\footnotesize (9)} & {\footnotesize (10)} \\ \hline
\multicolumn{10}{c}{Cluster LCBGs}\\ \hline 
1081 & J045403.40-025920.3 & 73.51414576 & -2.98898847 & 23.48$\pm$0.02 & 23.90$\pm$0.02 & 22.57$\pm$0.02 & 22.06$\pm$0.02 & 21.96$\pm$0.02 & 20.04$\pm$0.06 \\
1093 & J045405.10-025939.1 & 73.52125964 & -2.99419854 & 20.21$\pm$0.02 & 20.80$\pm$0.02 & 19.86$\pm$0.02 & 19.39$\pm$0.02 & 19.19$\pm$0.02 & 17.31$\pm$0.02 \\
1252 & J045405.27-025950.0 & 73.52195174 & -2.99721636 & 22.97$\pm$0.02 & 23.11$\pm$0.02 & 21.63$\pm$0.02 & 20.96$\pm$0.02 & 20.69$\pm$0.02 & 18.56$\pm$0.02 \\
1348 & J045406.22-030020.3 & 73.52593350 & -3.00565726 & 23.28$\pm$0.02 & 23.88$\pm$0.02 & 22.88$\pm$0.02 & 22.50$\pm$0.02 & 22.39$\pm$0.03 & {\bf 23.71}$\pm$1.9 \\
1954 & J045412.07-030201.2 & 73.55029362 & -3.03367338 & 23.26$\pm$0.02 & 23.61$\pm$0.02 & 22.19$\pm$0.02 & 21.72$\pm$0.02 & 21.40$\pm$0.02 & 19.75$\pm$0.05 \\
1968 & J045412.68-030054.2 & 73.55284217 & -3.01505968 & 21.80$\pm$0.02 & 22.35$\pm$0.02 & 21.23$\pm$0.02 & 20.73$\pm$0.02 & 20.65$\pm$0.02 & 19.02$\pm$0.03 \\
2084 & J045413.51-030148.4 & 73.55633494 & -3.03015174 & 22.97$\pm$0.02 & 23.28$\pm$0.02 & 22.07$\pm$0.02 & 21.45$\pm$0.02 & 21.25$\pm$0.02 & 18.99$\pm$0.03 \\
2224 & J045414.98-030424.4 & 73.56242894 & -3.07345646 & 22.73$\pm$0.02 & 23.05$\pm$0.02 & 21.64$\pm$0.02 & 20.98$\pm$0.02 & 20.94$\pm$0.02 & 18.85$\pm$0.02 \\
2300 & J045415.18-030331.9 & 73.56325367 & -3.05887289 & 22.32$\pm$0.02 & 22.97$\pm$0.02 & 22.11$\pm$0.02 & 21.74$\pm$0.02 & 21.88$\pm$0.02 & 19.78$\pm$0.05 \\
2932 & J045421.25-030325.9 & 73.58855821 & -3.05721700 & 23.09$\pm$0.02 & 23.62$\pm$0.02 & 22.65$\pm$0.02 & 22.32$\pm$0.02 & 22.39$\pm$0.03 & {\bf 20.58}$\pm$0.11 \\
664 & J045358.78-025857.2 & 73.49492161 & -2.98256249 & 22.99$\pm$0.02 & 23.39$\pm$0.02 & 21.81$\pm$0.02 & 21.15$\pm$0.02 & 21.08$\pm$0.02 & 19.05$\pm$0.03 \\
910 & J045401.42-030125.6 & 73.50592319 & -3.02380154 & 22.48$\pm$0.02 & 23.02$\pm$0.02 & 21.95$\pm$0.02 & 21.51$\pm$0.02 & 21.44$\pm$0.02 & 19.85$\pm$0.06 \\
947 & J045401.78-030054.4 & 73.50739764 & -3.01511665 & 22.00$\pm$0.02 & 22.58$\pm$0.02 & 21.91$\pm$0.02 & 21.45$\pm$0.02 & 21.32$\pm$0.02 & 19.54$\pm$0.04 \\
950 & J045401.86-030029.2 & 73.50775986 & -3.00812351 & 21.18$\pm$0.02 & 21.80$\pm$0.02 & 20.94$\pm$0.02 & 20.48$\pm$0.02 & 20.43$\pm$0.02 & 18.93$\pm$0.03 \\
2947 & J045421.31-030230.0 & 73.58880494 & -3.04166653 & 24.82$\pm$0.05 & 24.72$\pm$0.03 & 23.05$\pm$0.02 & 22.66$\pm$0.02 & 22.07$\pm$0.03 & {\bf 20.75}$\pm$0.13 \\
3201 & J045424.11-030348.0 & 73.60043585 & -3.06333827 & 23.03$\pm$0.02 & 23.26$\pm$0.02 & 21.82$\pm$0.02 & 21.18$\pm$0.02 & 20.92$\pm$0.02 & 18.74$\pm$0.02 \\ \hline
\multicolumn{10}{c}{Cluster Blue Galaxies}\\ \hline 
1118 & J045404.50-030013.6 & 73.51873535 & -3.00379952 & 21.90$\pm$0.02 & 22.21$\pm$0.02 & 20.68$\pm$0.02 & 19.94$\pm$0.02 & 19.62$\pm$0.02 & 17.24$\pm$0.02 \\
2312 & J045415.86-030356.1 & 73.56606808 & -3.06559628 & 22.17$\pm$0.02 & 22.56$\pm$0.02 & 21.27$\pm$0.02 & 20.69$\pm$0.02 & 20.57$\pm$0.02 & 18.52$\pm$0.02 \\
629 & J045358.36-030126.6 & 73.49317917 & -3.02408574 & 24.12$\pm$0.03 & 24.12$\pm$0.02 & 22.38$\pm$0.02 & 21.66$\pm$0.02 & 21.43$\pm$0.02 & 19.08$\pm$0.03 \\
732 & J045359.56-025803.5 & 73.49816162 & -2.96765325 & 22.82$\pm$0.02 & 23.06$\pm$0.02 & 21.48$\pm$0.02 & 20.72$\pm$0.02 & 20.48$\pm$0.02 & 18.32$\pm$0.02 \\
814 & J045400.13-030207.8 & 73.50053598 & -3.03551494 & 24.15$\pm$0.03 & 24.74$\pm$0.03 & 23.55$\pm$0.02 & 22.93$\pm$0.03 & 22.76$\pm$0.03 & {\bf 21.80}$\pm$0.33 \\
925 & J045401.41-025859.8 & 73.50586864 & -2.98329587 & 23.24$\pm$0.02 & 23.66$\pm$0.02 & 22.47$\pm$0.02 & 22.02$\pm$0.02 & 21.83$\pm$0.02 & {\bf 20.61}$\pm$0.11 \\
3861 & J045428.96-030506.8 & 73.62064367 & -3.08519410 & 22.90$\pm$0.02 & 23.47$\pm$0.02 & 22.32$\pm$0.02 & 21.72$\pm$0.02 & 21.78$\pm$0.02 & {\bf 20.62}$\pm$0.11 \\
\hline
\end{tabular} 
}
\footnotesize
\begin{flushleft}
Note. -- (1) Object identification number (2) Identification in the WLTV survey (3) Right Ascension (4) Declination (5) $U$-band magnitude and error (6) $B$-band magnitude and error (7) $R$-band magnitude and error (8) $I$-band magnitude and error (9) $z$-band magnitude and error (10) $K$-band magnitude and error. Objects detected below the $K$-band limiting magnitude have corresponding photometry in bold.
\end{flushleft}
\end{table*}

\section{Observations \& Data Analysis}  \label{sec:Observations and Data Analysis}
We describe the observations forming the basis of the present study in our previous paper \citep{2016ApJ...817...87C}. Here, we summarize these observations and describe the additional observations and analysis in the following sections. 

\subsection{Photometric Data and Analysis}
\subsubsection{Observations}

We obtained deep optical images in the $UBRIz$ passbands between 1999 October and 2004 June with the WIYN\footnote{The WIYN Observatory is a joint astronomical facility of the University of Wisconsin-Madison,
  Indiana University, Yale University, and the National Optical
  Astronomy Observatory.} 3.5~m telescope's Mini-Mosaic Camera.
  The final reduced data has a $9\farcm6\times9\farcm6$ field of view, $0\farcs14$ per pixel sampling, and $0\farcs74$ FWHM seeing for the final combined R-band image. Further details of the observations and data reductions appear
in \cite{2009ApJ...690.1158C,2016ApJ...817...87C}.

Deep NIR $K$-band observations\footnote{\url {http://www.astro.caltech.edu/clusters/}}
were supplied by \cite{2007ApJ...671.1503M} based on observations taken with the WIRC camera on the Hale 5~m telescope at Palomar
Observatory\footnote{The Palomar Observatory is an astronomical
  facility located in north San Diego County, California, USA.}. The
archive image has a $8\farcm7 \times 8\farcm7$ field of view, $0\farcs25$ per pixel
sampling, and $0\farcs97$ FWHM median seeing.   
Further details of the observations are 
provided in the original paper \citep[see][]{2007ApJ...671.1503M}.  

\subsubsection{Photometry \& Size Analysis}

Following \cite{2009ApJ...690.1158C, 2016ApJ...817...87C}, prior to photometric analysis we aligned all images and matched the seeing to the worst band (i.e., the $K$ band) by convolving the images with a 2-D Gaussian function. This  enabled our photometric apertures to sample the same light profiles from the six bands ($UBRIzK$). We used the SExtractor software package \citep{1996A&AS..117..393B} in two-image mode with source detection performed on the $R$ band image, our deepest image, to produce matched aperture photometry in the different bands.  We used a photometric aperture diameter of 7.5~kpc at the redshift of the cluster to measure magnitudes for each cluster source. This size was chosen to provide consistency with the previous study done by \cite{2003ApJ...586L..45G}. For most of our sources this measured greater than 96\% of the total light.  Table \ref{tab:photometry} presents the optical and near-infrared photometry. The errors presented, and used for the fitting, are the Poisson errors combined (in quadrature) with an additional 2\% error measured from simulations of the observations \citep{2009ApJ...690.1158C}. 
 
As presented in \cite{2016ApJ...817...87C}, we measured the size of each star-forming cluster galaxy based on imaging of the cluster from the HST Advanced Camera for Surveys \cite[ACS;][]{2003SPIE.4854...81F}.  The ACS data (PI: Ellis, Proposal ID: 9836) cover the entire field of view of the WIYN data in the F814W band with typical exposure times of 2036~s \cite[see][]{2016ApJ...817...87C}.   For each galaxy, we measured the total flux via an iterative analysis of the curve of growth, then determined the half-light radius ($r_{e}$) following the methodology in \cite{2006ApJ...636L..13C}.  This method does not assume a specific profile shape and will measure accurate sizes for galaxies having various shapes and sizes. 

\subsection{Spectroscopic Data and  Analysis}  
\subsubsection{Observations}

We completed spectroscopic observations in October 2005 with the Deep Imaging Multi-Object Spectrograph \citep[DEIMOS;][]{2003SPIE.4841.1657F} on the Keck~II telescope.  The multi-slit observations employed a slit width of $1\farcs0$ with the 900 l~mm$^{-1}$ grating to give a dispersion of 0.44~\AA\ pix$^{-1}$ and corresponding spectral resolution of  $2.85$~\AA\ (FWHM) across the observed wavelength range of 5000--8400~\AA.  We observed objects using four different slitmasks during the run, preferentially targeting potential LCBGs and other star-forming galaxies in the cluster.  Full details of the observations and data reductions appear in \cite{2011ApJ...741...98C}.

\subsubsection{Velocity Dispersion \& Dynamical Mass}
As described in \cite{2016ApJ...817...87C},  the internal gas velocity dispersion ($\sigma_v$) was calculated for each cluster galaxy by fitting Gaussian profiles to the [\Oii]~$\lambda$3727, H$\beta$, and [\Oiii]~$\lambda$5007 emission lines. For the [\Oii]~$\lambda$3727 doublet we fit a double Gaussian with fixed separation of 2.7 \AA\ in  the rest frame, constraining the peaks to have the same width while allowing their relative amplitudes to vary. We inspected all fits visually to confirm quality.  We estimated the instrumental spectral broadening from sky lines observed near our lines of interest, and subtracted the instrumental dispersion in quadrature from the measured velocity dispersion to correct for this effect.  At least one mask was observed under exceptional seeing conditions ($\sim0\farcs8$), and we corrected the instrumental velocity dispersion to the appropriate spectral resolution based on the seeing disc rather than the slit width for that set of spectra.  We determined the final velocity dispersion for each source by computing a weighted average of the measurement of the three different emission lines.  The smallest velocity dispersion we can safely recover is $10~\kms$ \citep{2016ApJ...817...87C}, and the typically error on each measurement is less than $10\%$.  Emission line signal to noise was typically greater than 10. 

We next converted the velocity dispersion into a dynamical mass via Equation (2) of \cite{2016ApJ...817...87C} viz. 
\begin{equation}
\Mdyn = \frac{3 c_2}{G} \sigma_v^2 r_{e}
\end{equation}
where $c_{2}$ = 1.6 \cite[see][]{1992ApJ...399..462B,1997ApJ...489..543P}. This value was chosen to match earlier works and assumes that the galaxies are virialised.  Following \cite{2003ApJ...586L..45G}, we applied a correction factor of 1.3 to the measured velocity dispersions to account for using nebular emission lines instead of absorption features in measuring the galaxy masses. We expect there are systematics in this mass formulation (e.g., in the specific value of $c_{2}$ or the correction factor for a given object). Our goal was to adopt a formalism  mirroring the analysis of the field sample and thereby eliminate systematics in the comparison of these two populations due to differing dynamical assumptions.

\subsection{Sample Selection} 

\begin{figure}
\centering
\includegraphics[width=0.5\textwidth]{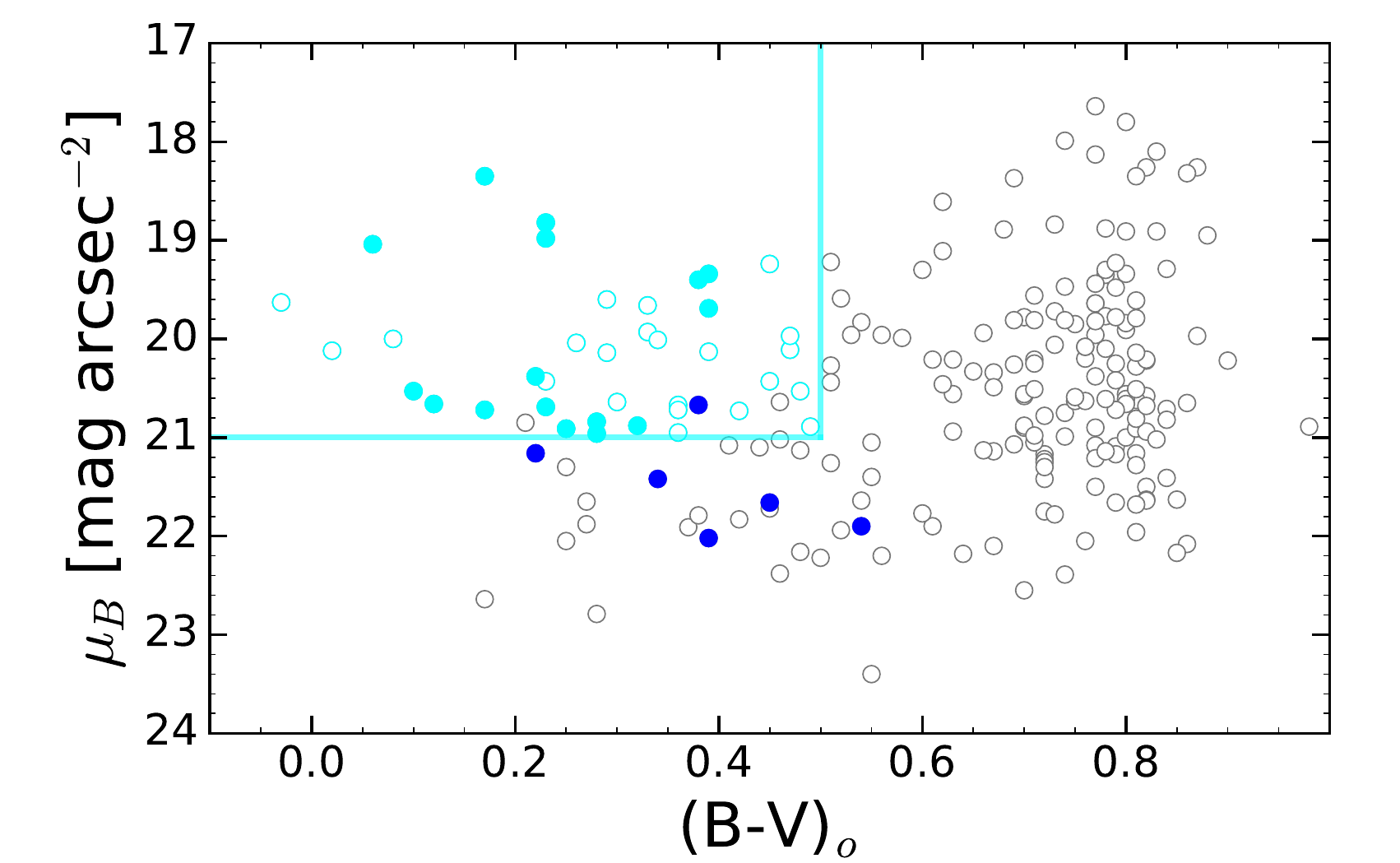}
\caption{\label{fig:sbbv} Selection window for LCBGs in terms of rest-frame color and mean rest-frame surface brightness within the half-light radius. Open circles represent spectroscopically-confirmed cluster members in MS\,0451-03; teal objects represent LCBGs. Filled circles indicate those sources used in this work, including a subset of blue galaxies (blue circles) spectroscopically confirmed as cluster members. Teal lines delineate color and surface-brightness cuts used to define LCBGs in this study; unmarked objects within this region are insufficiently bright to quality as LCBGs ($M_B > -18.5$).}
\end{figure}

The sample of cluster star-forming galaxies studied here is defined and classified in \cite{2016ApJ...817...87C}. From our sample, we selected a subsample of 23 galaxies, consisting of 16 cluster LCBGs and 7 cluster blue galaxies. These are the 23 star-forming galaxies (i.e., not on the red sequence) for which we have Keck spectra to measure emission line-widths and strengths, and which are also confirmed cluster members (previous spectroscopic surveys identified additional cluster members but either focus on the red-sequence, offer insufficient spectral resolution to measure line-widths, or have not shared the reduced spectra). Table~\ref{tab:params01} presents the final subsample including both classes of galaxies, and Figure~\ref{fig:sbbv} shows how we defined our sources based on color and surface brightness. Following \cite{2006ApJ...636L..13C}, we defined LCBGs as having $B-V < 0.5$, $\mu_B < 21$ mag arcsec$^{-2}$, and $M_B < $-18.5. We have  measured both stellar and dynamical masses of galaxies for the full subsample of 23 targets. 

In this work, we primarily aim to compare the measured stellar and dynamical masses of our  cluster LCBGs to the corresponding field LCBG sample of \cite{2003ApJ...586L..45G}. Their sample was defined and classified in \cite{1997ApJ...489..543P,1997ApJ...489..559G}. We note that the criteria they employed to define their field LCBG sample (half-light radius $r_{e} \leq 0\farcs5$, magnitude $I_{814}\leq23.74$, and surface brightness $\mu_{I_{814}}<22.2~\rm mag~arcsec^{-2}$)  correspond closely to the definition employed herein  when translated into absolute properties \cite[see][]{2016ApJ...817...87C}. To ensure a valid comparison, we consider only the 15 sources from the field sample lying within the redshift range $0.3 < z < 0.7$. 

\begin{figure*}
\begin{center} 
\includegraphics[width=0.95\textwidth]{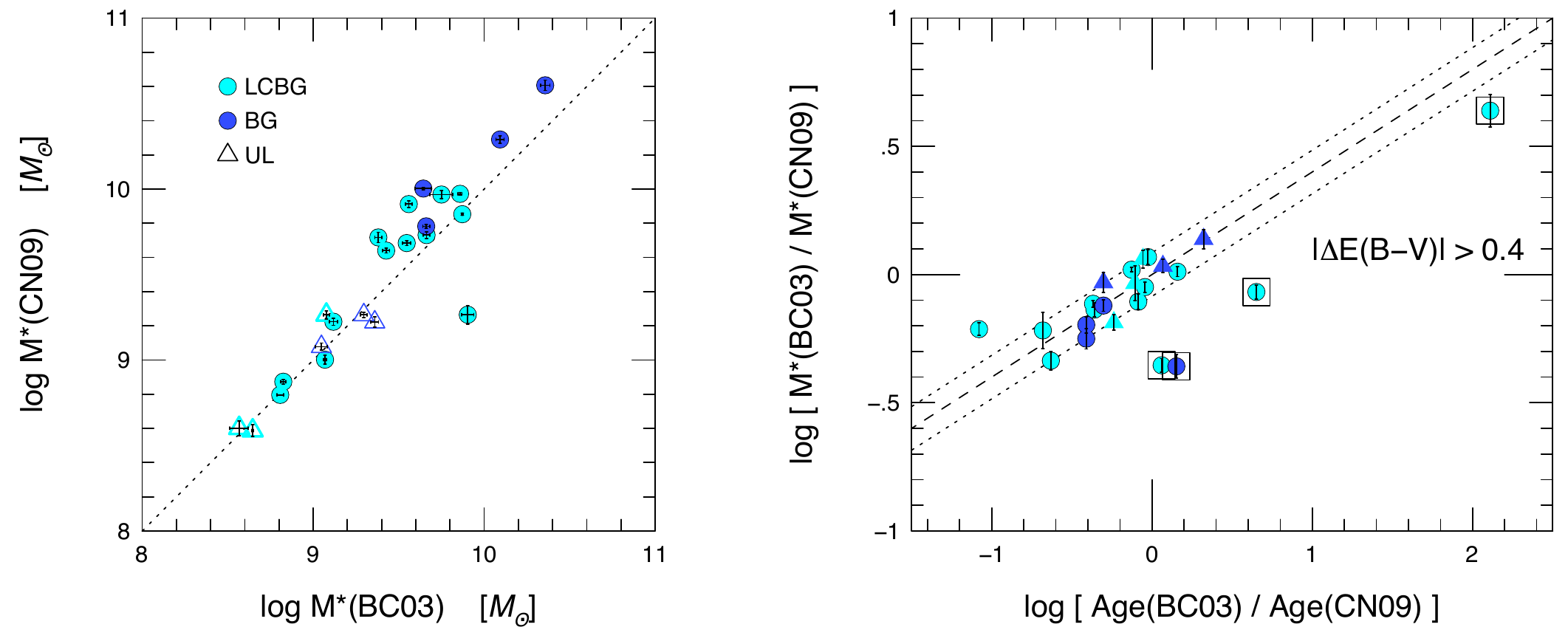}
\end{center} 
\caption{\label{fig:best-fit-masses}(Left)~Comparison of best-fit stellar masses measured by  BC03 and CN09 models. Model masses correlate strongly but with a scatter larger than the formal errors.  Masses derived from the CN09 models are systematically higher at greater mass.  (Right)~Ratio of the best-fit mass by the two models as a function of the best-fit age. The dashed and dotted lines are a representative fit, consistent with $d\log M/d\log A  = 0.4$ and a scatter of about 0.09 dex (25\%). This scatter is consistent with the formal errors. Galaxies with upper limits on their stellar masses are shown in triangular points and it is donated as UL in both left and right panels. The square markers in the right-hand panel indicate galaxies that show an extinction significantly off compared to the rest of the sample.}
\end{figure*}

\section{Stellar Mass Estimates} 
\label{sec:Stellar Mass Estimates}

The most common and reliable method for estimating the stellar masses of galaxies is to fit a model spectral energy distribution (SED) to the observed flux \cite[e.g.][]{2004ApJ...616L.103D}. Under this method, the mix of star types, star ages, and interstellar extinction within the galaxy define the shape of the galaxy SED, and the normalization factor required to reproduce the observed absolute flux determines the total stellar mass of the galaxy. Although this technique is usable across various optical and infrared passbands, it works particularly well in the near-IR $K$ passband because this spectral region is relatively unaffected by dust extinction \citep{1998MNRAS.297L..23K} and is most sensitive to the lower-mass stars that constitute the bulk of the mass within a galaxy \citep{2000ApJ...536L..77B,2003ApJ...586L..45G}. Given the relative photometric errors in this data-set, the $K$ band (rest-frame wavelength $\sim1.4\mu$m) plays a relatively modest role in constraining the models, but still provides a long wavelength anchor.

\subsection{SED fitting} 
\label{subsec:SED-fitting} 

Rather than relying on a single SED model to compare with our galaxy spectra, we derived results using two fundamentally distinct SED models to understand the effect of the differing approaches.  Our two selected models were the public SPS GALAXEV code of \citet[][hereafter BC03]{2003MNRAS.344.1000B} and the more recent flexible stellar population synthesis (FSPS) code developed and publicly released by \citet[][henceforth CN09]{2009ApJ...699..486C}.

GALAXEV features high spectral resolution (3\AA) and FSPS is known to feature an easy handling of computation of SED models. CN09 SPS code has an improved treatment of TP-AGB phase since \cite{2009ApJ...699..486C} modified the isochrone synthesis of Padova while BC03 SPS code has not.

For internal consistency and comparison to previous modeling of field LCBGs \citep{1997ApJ...489..559G} we chose a single initial mass function (IMF) for both models. Our choice  of the Salpeter \citep{1955ApJ...121..161S} IMF may not be the most accurate for star-forming galaxies \citep[e.g.,][]{2014ApJ...796...75C} but it is useful as a likely upper limit to the inferred stellar mass for comparison to our dynamical mass estimates.  For reference, adopting a Chabrier \citep{2003PASP..115..763C} IMF would yield masses $\sim$60\% of values estimated here \citep{2012Natur.484..485C}.

\subsubsection{Ingredients for SPS Codes}
 
Table~\ref{tab:params Grid} summarizes the parameters controlling our SED library, including the IMF type, star formation history (simple stellar population, burst, exponential, or constant SFR), age, metallicity,  and the dust extinction model and intensity.  We based our choice of input components for computing the two SED model libraries on a similar grid of ingredients by \cite{2003ApJ...586L..45G}. The selected parameters may not reflect the true conditions in these galaxies; for example, the well-known degeneracy between age and metallicity is no doubt manifest in our models and fitting results. Our parameter choices may thus introduce systematic differences to the bona fide stellar masses, as we explore below.

\subsubsection{Stellar Mass Measurements} 

Given the library of model SEDs that we generated with the BC03 and CN09 SPS codes across a wide range of metallicity, age, star formation history, and dust content, we employed an SED-fitting technique to compare these models to the optical-NIR photometric data \cite[see][]{2000ApJ...536L..77B,2003ApJ...586L..45G}. This enabled  us to infer the stellar masses and other physical properties of each galaxy, and  how systematic differences in these properties between the two sets of models  correlate.

Following a standard SED-fitting approach described in the literature \cite[see][]{2001A&A...368...74M,2013MNRAS.435...87M}, we employed the following goodness-of-fit metric:

\begin{equation} 
\rm \chisq = \sum_{i=1}^{n_{filters}}\left(\frac{F_{obs,i} - {\it N}\times 
F_{model,i}}{\sigma_{i}}\right)^{2}
\label{eq:chi2}
\end{equation}
where $\rm F_{obs,i}$, $ \rm F_{model,i}$, $\rm \sigma_{i}$, and $N$ are the observed (apparent) flux, model luminosity for M$_{\rm m}$ solar masses in stars, photometric error in the $i^{\rm th}$ passband, and constant of normalization, respectively. The value of  $\chisq$ is determined by summing over all available $UBRIzK$ photometric passbands.

To compare the models with our photometric results, we converted the model fluxes to broad-band magnitudes by shifting each spectrum to the redshift of the galaxy and convolving with the suitable response function. We compared the observed SED of each galaxy to a grid of 24,480 (BC03) and 47,520 (CN09) SED models, thus identifying the best-fitting SED as that associated with the minimum $\chisq$ value.  

For each galaxy, we recorded the SED best-fit model parameters, minimum $\chisq$ value, and the constant of normalization $N$.  Given the luminosity distance $\rm D_{L}$ at each object's redshift $z$, we estimated the stellar mass from the SED best-fit model via 
\begin{equation}
\label{eq:Massg}
\rm M_{*} = 4\pi D_{L}^{2} \times {\it N} \times M_{m}.
\end{equation}
where M$_{\rm m}$ is the model mass generated by the SPS for each SED.  We also derived the age of the stellar population, the extinction, and the metallicity from the SED best-fit parameters. The best-fitting model parameters for each target appear in Table~\ref{tab:params003} of the Appendix.

\begin{table*}
\centering
\caption{Measured and Derived Properties of Cluster Star-forming Galaxies.} 
\label{tab:params01} 
\begin{tabular}{r r r r r r r r c} 
\hline \hline 
\multicolumn{1}{c}{Obj} & 
\multicolumn{1}{c}{$z$}  & 
\multicolumn{1}{c}{$\Mast$(BC03)}  & 
\multicolumn{1}{c}{$\Mast$(CN09)}  & 
\multicolumn{1}{c}{r$_{e}$} & 
\multicolumn{1}{c}{$\sigma$} & 	
\multicolumn{1}{c}{$\Mdyn$} & 
\multicolumn{1}{c}{SFR (\Oii)} &
\multicolumn{1}{c}{($\Mdyn/\Mast$(BC03))} \\  
\multicolumn{1}{c}{(ID)}   &   & 
\multicolumn{1}{c}{($10^{9}\ \Msun$)} & 
\multicolumn{1}{c}{($10^{9}\ \Msun$)} & 
\multicolumn{1}{c}{(kpc)} & 
\multicolumn{1}{c}{($\kms$)} &  
\multicolumn{1}{c}{(10$^{9}~\Msun$)}  & 
\multicolumn{1}{c}{($\Msun$ yr$^{-1}$)}  & 
\multicolumn{1}{c}{} \\ 
\multicolumn{1}{c}{(1)}    & 
\multicolumn{1}{c}{(2)} & 
\multicolumn{1}{c}{(3)} & 
\multicolumn{1}{c}{(4)} & 
\multicolumn{1}{c}{(5)} & 
\multicolumn{1}{c}{(6)} & 
\multicolumn{1}{c}{(7)} & 
\multicolumn{1}{c}{(8)} &
\multicolumn{1}{c}{(9)} \\ \hline 
\multicolumn{9}{c}{Cluster LCBGs}\\ \hline 
1081 & 0.531 & 1.18 $\pm$ 0.03     & 1.14 $\pm$ 0.06    & 2.33 $\pm$ 0.47 & 72 $\pm$ 8 & 13.8 $\pm$ 3.2 & 1.42 $\pm$ 0.3 & 11.7 \\
1093 & 0.527 & 7.2 $\pm$ 0.9       & 8.9 $\pm$ 0.5      & 6.84 $\pm$ 0.18 & 49 $\pm$ 3 & 18.4  $\pm$ 1.3 & 1.53 $\pm$ 0.5 & 2.6 \\
1252 & 0.539 & 7.9 $\pm$ 0.2       & 9.3 $\pm$ 0.2      & 2.40 $\pm$ 0.12 & 74  $\pm$  26 & 14.8 $\pm$ 5.3 & 1.26 $\pm$ 0.1 & 1.9 \\
1348 & 0.531 &$<$0.58 $\pm$ 0.05   &$<$0.61 $\pm$ 0.04  & 1.63 $\pm$ 0.04 & 50  $\pm$  24 & 4.6  $\pm$  2.2 & 0.97  $\pm$ 0.2 & 7.9 \\
1954 & 0.528 & 4.7 $\pm$ 0.7       & 2.8 $\pm$ 0.2      & 2.36 $\pm$ 0.36 & 42  $\pm$  3 & 4.7  $\pm$  0.8 & 1.27  $\pm$  0.1 & 1.0 \\ 
1968 & 0.544 & 3.3 $\pm$ 0.1       & 4.2 $\pm$ 0.1      & 1.87 $\pm$ 0.10 & 47  $\pm$  28 & 4.8  $\pm$  2.8 & 0.84  $\pm$  0.0 & 1.5 \\
2084 & 0.548 & 4.0 $\pm$ 0.2       & 4.9 $\pm$ 0.2      & 3.12 $\pm$ 0.41 & 77  $\pm$  28 & 20.7  $\pm$  8.0 & 1.35  $\pm$  0.1 & 5.2 \\
2224 & 0.530 & 4.1 $\pm$ 0.2       & 7.0 $\pm$ 0.4      & 3.40 $\pm$ 0.33 & 80  $\pm$  4 & 24.7  $\pm$  2.8 & 1.53  $\pm$  0.1 & 6.0 \\
2300 & 0.535 & 0.65 $\pm$ 0.03     & 0.624 $\pm$ 0.002  & 1.85 $\pm$ 0.16 & 44  $\pm$  2 & 4.2  $\pm$  0.4 & 2.66  $\pm$  0.2 & 6.5 \\
2932 & 0.542 &$<$0.440 $\pm$ 0.005 &$<$0.48 $\pm$ 0.03  & 2.26 $\pm$ 0.36 &  $<$ 10  & <0.2  & 0.75  $\pm$  0.1 & 0.5 \\
664	 & 0.538 & 3.7 $\pm$ 0.2       & 4.9 $\pm$ 0.2      & 3.82 $\pm$ 1.08 & 26  $\pm$  2 & 3.0  $\pm$  0.9 & 1.38  $\pm$  0.2  & 0.8 \\
910  & 0.540 & 1.23 $\pm$ 0.07     & 1.53 $\pm$ 0.08    & 2.10 $\pm$ 0.18 & 37  $\pm$  3 & 3.4  $\pm$  0.5 & 2.88  $\pm$  0.2 & 2.8   \\
947	 & 0.532 & 0.68 $\pm$ 0.02     & 0.78 $\pm$ 0.02    & 2.78 $\pm$ 0.02 & 66  $\pm$  7 & 13.8  $\pm$  1.6 & 2.71  $\pm$  0.2 & 20.3 \\  
950	 & 0.511 & 2.4 $\pm$ 0.1       & 3.8 $\pm$ 0.4      & 1.31 $\pm$ 0.03 & 53  $\pm$  3 & 4.2  $\pm$  0.3 & 5.26  $\pm$  0.1 & 1.8 \\
2947 & 0.511 &$<$1.19 $\pm$ 0.05   &$<$1.5 $\pm$ 0.1    & 2.29 $\pm$ 0.55 & 45  $\pm$  4 & 5.2  $\pm$  1.4 & 0.51  $\pm$  0.1 & 4.4 \\ 
3201 & 0.506 & 7.45 $\pm$ 0.08     & 7.2 $\pm$ 0.1      & 2.08 $\pm$ 0.12 & 62  $\pm$  7 & 9.2  $\pm$  1.2 & 1.05  $\pm$  1.0 & 1.2 \\ \hline 
\multicolumn{9}{c}{Cluster Blue Galaxies}\\ \hline 
1118 & 0.532 &22.2 $\pm$ 1.5       & 30 $\pm$ 3         & 7.39 $\pm$ 0.38 &  54.7 $\pm$ 11  & 24.7 $\pm$ 5 & 2.31 $\pm$ 0.1 & 1.1  \\ 
2312 & 0.526 & 5.5 $\pm$ 0.5       &  9.8 $\pm$ 0.1     & 5.05 $\pm$ 0.48 &  45.5 $\pm$ 3  & 11.6 $\pm$ 1.3 & 3.12 $\pm$ 0.1 & 2.1  \\ 
629  & 0.537 & 4.8 $\pm$ 0.2       &  5.8 $\pm$ 0.1     & 4.43 $\pm$ 0.10 &  ...  & ... & 0.88 $\pm$ 0.1 & ...  \\
732  & 0.551 & 12.8 $\pm$ 0.7      & 14 $\pm$ 1         & 5.65 $\pm$ 0.55 &  91.8 $\pm$ 22  & 53 $\pm$ 1.16 & $ 1.12 \pm$ 0.1 & 4.1  \\
814  & 0.548 &$<$1.1$ \pm$ 0.1     &$<$1.09 $\pm$ 0.06  & 1.63 $\pm$ 0.39 &  <10  & <0.2 & 0.65 $\pm$ 0.2 & 0.2   \\
925  & 0.531 &$<$1.9$ \pm$ 0.1     &$<$1.1 $\pm$ 0.1    & 2.96 $\pm$ 1.42 &  44.7 $\pm$ 15  & 6.6 $\pm$ 4 & 0.68 $\pm$ 0.1 & 3.5   \\
3861 & 0.509 &$<$2.0$ \pm$ 0.1     &$<$1.80 $\pm$ 0.07  & 4.24 $\pm$ 1.06 &  24.9 $\pm$ 6.3 & 2.9 $\pm$ 1 & 0.73 $\pm$ 0.1 & 1.4 \\ \hline
\end{tabular} 
\begin{flushleft}
Note. -- (1)~Object identification number; (2)~redshift; (3)~mean and standard deviation of stellar mass from MC simulations using BC09 models; (4)~mean and standard deviation of stellar mass from MC simulations using CN09 models; (5)~half-light radius; (6)~velocity dispersion estimated from [\Oii]~$\lambda$3727; (7)~dynamical mass; (8)~SFR derived from [\Oii]~$\lambda$3727, (9)~$\Mdyn/\Mast$(BC03)). Half-light radius, velocity dispersion, dynamical masses, and the star formation rates measurements from \protect\cite{2016ApJ...817...87C} are described in Section \ref{sec:Observations and Data Analysis}.
\end{flushleft}
\end{table*}

\begin{table*}
\centering
\caption{Grid of parameters used to compute of SEDs from SPS models of BC03 and CN09.} 
\label{tab:params Grid} 
\begin{center}
\begin{tabular}{l l l l l} \hline \hline 
Parameter  & Units			 & Parameter range (BC03) 			& Parameter range (CN09) & Note \\ \hline 
Age 		  & Gyr  			 & 0.001 up to 14.0 (via $\Delta t$)					& 0.0003 up to 14.0 (via $\Delta t$) & (1) \\ 
SFR [SSP]   &  	$\cdots$		 & SSP 			& SSP & (2)\\ 
SFR [B.~(length)]   & Gyr 			 & 0.5, 1.0, 1.5, 2.0 			& 0.5, 1.0, 1.5, 2.0 & (3)\\ 
SFR [E. ($\tau$)]  & Gyr 			 & 0.5, 1.0, 1.5, 2.0 					& 0.5, 1.0, 1.5, 2.0 & (4) \\ 
SFR [C.]  &$\cdots$ 		 & $\cdots$					& $\cdots$ & (5)\\
Metallicity  &$Z_\odot$		 	& 0.005, 0.02, 0.2, 0.4, 1, 2.5 			&  0.025, 0.05, 0.2, 0.4, 1, 1.6\\
E(B-V)       &mag 			& 0, 0.1, 0.25, 0.5, 1 				&  0, 0.1, 0.25, 0.5, 1 & (6)\\
Extinction law &	$\cdots$		 	& MW, CZ					  	& MW, CZ & (7)\\ 
IMF 		  &		$\cdots$		 & Salpeter     		& Salpeter  & (8)\\ \hline 
\end{tabular} 
\begin{flushleft}
Notes -- (1)~The age increment ($\Delta t$) is 0.0001~Gyr; (2)~Simple stellar population; (3) Burst model (B) with indicated duration (length) in Gyr; (4)~Exponentially-decaying star formation model (E) with the indicated time constant $\tau$ in Gyr; (5)~Constant star formation rate; (6)~Intrinsic extinction amount; (7)~Internal extinction law options are ``CZ'' for \cite{2000ApJ...533..682C} and ``MW'' for Milky Way. (8)~Assumed IMF. 

\end{flushleft}
\end{center}
\end{table*}

\subsection{Error Estimates} 

We estimated the uncertainty on the measured stellar masses through Monte Carlo (MC) simulations. In this method, we re-sampled the colours of each galaxy in accordance with the photometric errors, assuming those errors have a Gaussian distribution with the stated standard deviation. We then determined the SED best-fit stellar masses following the same procedure described in Section~\ref{subsec:SED-fitting}. 

We conducted 150 trials and recorded the mean stellar mass ($\langle\Mast\rangle$) and its standard deviation for each galaxy.  We adopt these standard deviations derived from the Monte Carlo simulations as our best estimate of the uncertainty in the stellar masses. The  average uncertainty for the measured stellar masses are $\sim0.27$~dex for BC03 and $\sim0.31$~dex for CN09. Table~\ref{tab:params01} presents the mean and standard deviation of stellar masses derived  through the MC simulations for BC03 and CN09 models.

The estimated uncertainties for our stellar mass measurements appear consistent with those reported by \cite{2003ApJ...586L..45G}, who found typical uncertainties of 0.36 dex in their full sample of field LCBGs.  Our typical uncertainties are comparable to both the $\sim0.3$~dex uncertainties found by \cite{2012MNRAS.420.1481V} for cluster galaxies and to the $\sim0.3$~dex uncertainties derived for distant field galaxies by \cite{2000ApJ...536L..77B}.

Additional systematic and random effects (e.g., photometric errors, incompleteness in the models, and more complex star formation histories) may contribute an additional $~0.2$~dex of uncertainty for which we have not accounted \cite[see][]{2001ApJ...559..620P,2005ApJ...625..621B}.  However, since similar errors may affect the field sample such systematics are of less significance for our relative comparison. 

\subsection{Caveats}

A significant fraction of the objects in our sample ($\sim40$\%) failed to yield satisfactory SED fits, producing reduced $\chisq > 10$.  Although this suggests a poor match between the best-fit models and our photometric measurements,  causes may include  (a)~underestimated random errors in the photometry, (b)~systematic errors in the photometry (background estimation), (c)~contamination from emission lines, (d)~incompleteness in the stellar libraries, and (e)~poor choices for model parameters ($Z$, dust extinction, and IMF).

Unfortunately, we lack $\chisq$ values for the fits that  \cite{2003ApJ...586L..45G} achieved with their field LCBG sample, which serves as our primary point of comparison.
Given that our stellar mass errors appear comparable to theirs, we decided not to reject any of our galaxies with fits that yield large $\chisq$ values.  However, to confirm data quality we visually inspected the fits to all our targets and defined a ``good'' fit as having achieved acceptable agreement in at least four filter passbands.  We retained the objects that did not meet this criterion but indicate their stellar masses as upper limits ($<$) in Table~\ref{tab:params01}. Figures~\ref{fig:A1},\ref{fig:A2},\ref{fig:A3},\ref{fig:A4} show all of the observed and best-fitting model SEDs for our sample.

\subsection{Comparison of Stellar Mass Measurements} 
\label{sec: Cluster Blue Star-forming Galaxies}

Here, we present  the estimated stellar masses derived from the two SPS models. The left-hand panel of Fig.~\ref{fig:best-fit-masses} compares the stellar masses obtained from the BC03 and CN09 models for all galaxies in our sample.  The best-fit line for the derived stellar masses is slightly steeper than the comparison line of equal mass.  The median stellar mass measured using the CN09 models is $\sim0.2$ dex higher than that measured using the BC03 models. We conclude that the two models give generally consistent results, echoing the similar finding made earlier by \cite{2010ApJ...712..833C} in comparing results from the FSPS and GALAXEV codes for massive red galaxies.  Both studies confirm the contention of \cite{2006A&A...458..717R} that stellar mass estimates derived from photometry are largely insensitive to the chosen SPS model.

The ratio of masses measured using the two different models does correlate with the age and extinction measured using each model as shown in the right-hand panel of Fig.~\ref{fig:best-fit-masses}.  The BC03 models tend to infer younger ages and smaller masses (and hence a lower M/L ratio).  For several galaxies, the difference in the adopted reddening is quite significant, and these objects clearly form a separate track in the figure (marked as boxes).   Independent measurements of the age or extinction for these galaxies would improve constraints on the stellar mass. 

Although the effect of the chosen SPS model may be small, it is (as demonstrated by both panels of Fig.~\ref{fig:best-fit-masses}) measurable.  Since the aim of this study is to compare properties of our cluster galaxies to those in the field sample of \cite{2003ApJ...586L..45G}, employing the same SPS model in both cases will eliminate a potential source of systematic error \cite[cf.][]{2010A&A...515A.101C}. We therefore adopt the BC03-derived stellar masses in our subsequent analysis for consistency with previous studies.

\begin{figure}
\centering
\includegraphics[width=.475\textwidth]{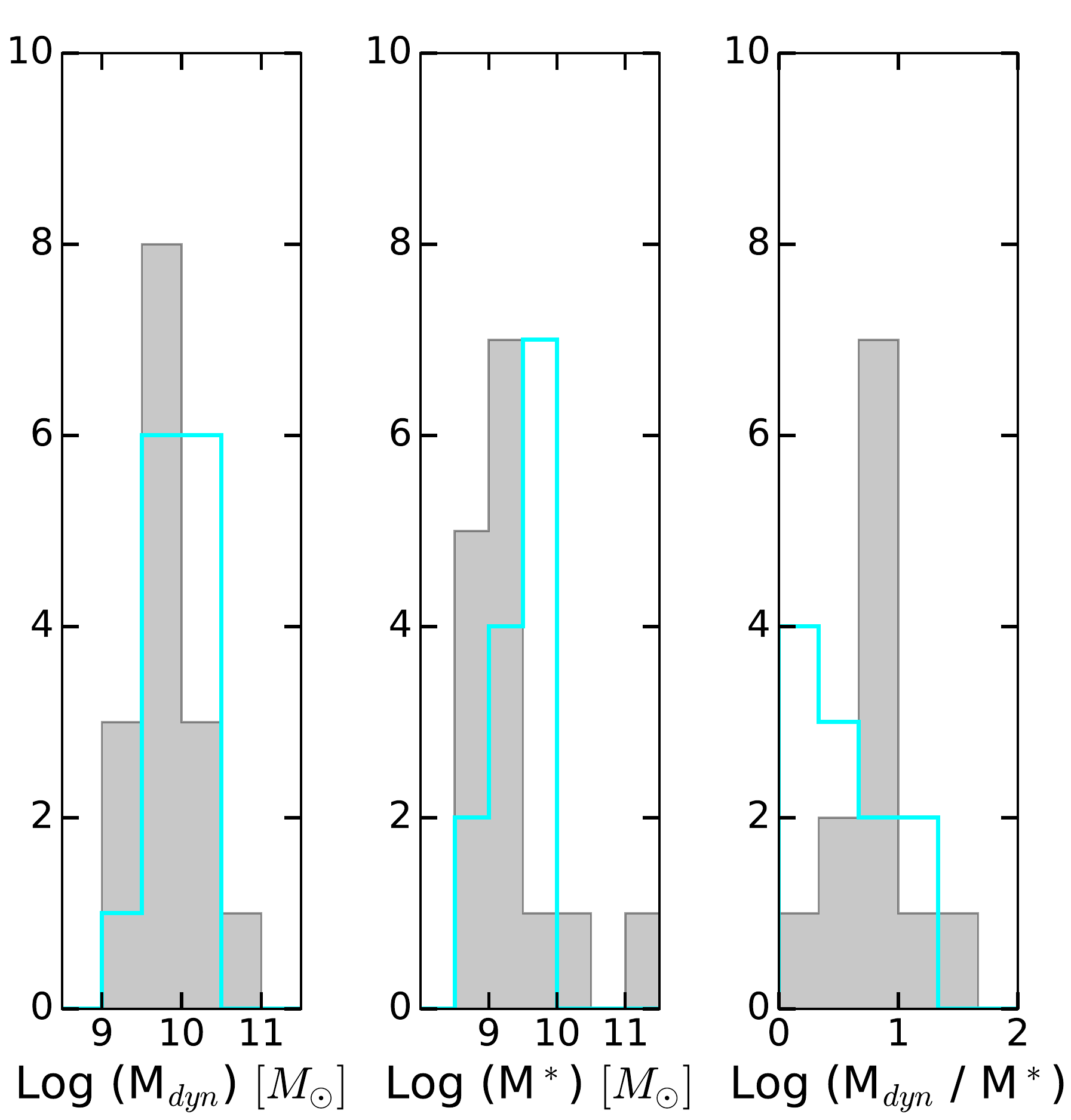}
\caption{{\it Left:} Histogram of cluster (teal) and field (grey) LCBGs dynamical masses.  {\it Center:}  Histogram of cluster and field LCBGs stellar masses using best-fit BC03 values. We find no strong evidence suggesting a significant difference between the two distributions as described in the text. {\it Right:} Histogram of the ratio of dynamical-to-stellar mass.   } 
\label{fig:Mstar-cluster-field}
\end{figure}

\section{Results} \label{sec: Results}

\subsection{Luminous Compact Blue Galaxies} 
\label{sec: Luminous Compact Blue Galaxies} 

\subsubsection{Mass Distributions}  

Here, we compare the measured stellar and dynamical masses of 16 cluster LCBGs to those of the field sample from \cite{2003ApJ...586L..45G}. Recall that the field sample we've adopted here includes 15 galaxies in the redshift range $0.3 < z < 0.7$ to facilitate a comparison to our cluster population. The field sample is complete in this redshift range to the luminosity limit of the LCBG class so that this selection, like the cluster sample, is volume limited. Hence, we can directly compare the mass and luminosity distributions of the cluster and field samples, modulo overall normalization (accounting for differences in total volume and space density between samples) which is not of interest here. The center panel in Figure~\ref{fig:Mstar-cluster-field} shows the distribution of stellar masses for the cluster and field LCBGs. The two distributions are broadly similar. The median stellar mass for the cluster LCBGs ($\Mratio \approx 9.40$) is similar to the field sample ($\Mratio \approx 9.76$). The cluster LCBGs have slightly lower masses when compared to their field counterparts ($\sim$ 0.36 dex) while the distribution of field LCBGs shows a high-mass tail that is absent in the cluster sample, albeit in a sample of limited size.

We performed a Kolmogorov-Smirnov (K-S) test to investigate the differences between the distributions of the two samples, with a corresponding $P_{K-S}$ of 0.11 on stellar masses (1-D test) and $P_{K-S}$ of 0.6 on the dynamical masses. The K-S test could not reject the null hypothesis that the two stellar masses and samples distributions are drawn from the same parent population (at 5\% significance).  Hence, we find no strong evidence suggesting  a significant difference between the stellar and dynamical mass distributions for the two samples (distributions are similar for the two environments, within the errors).

\begin{figure}
\includegraphics[width=.475\textwidth]{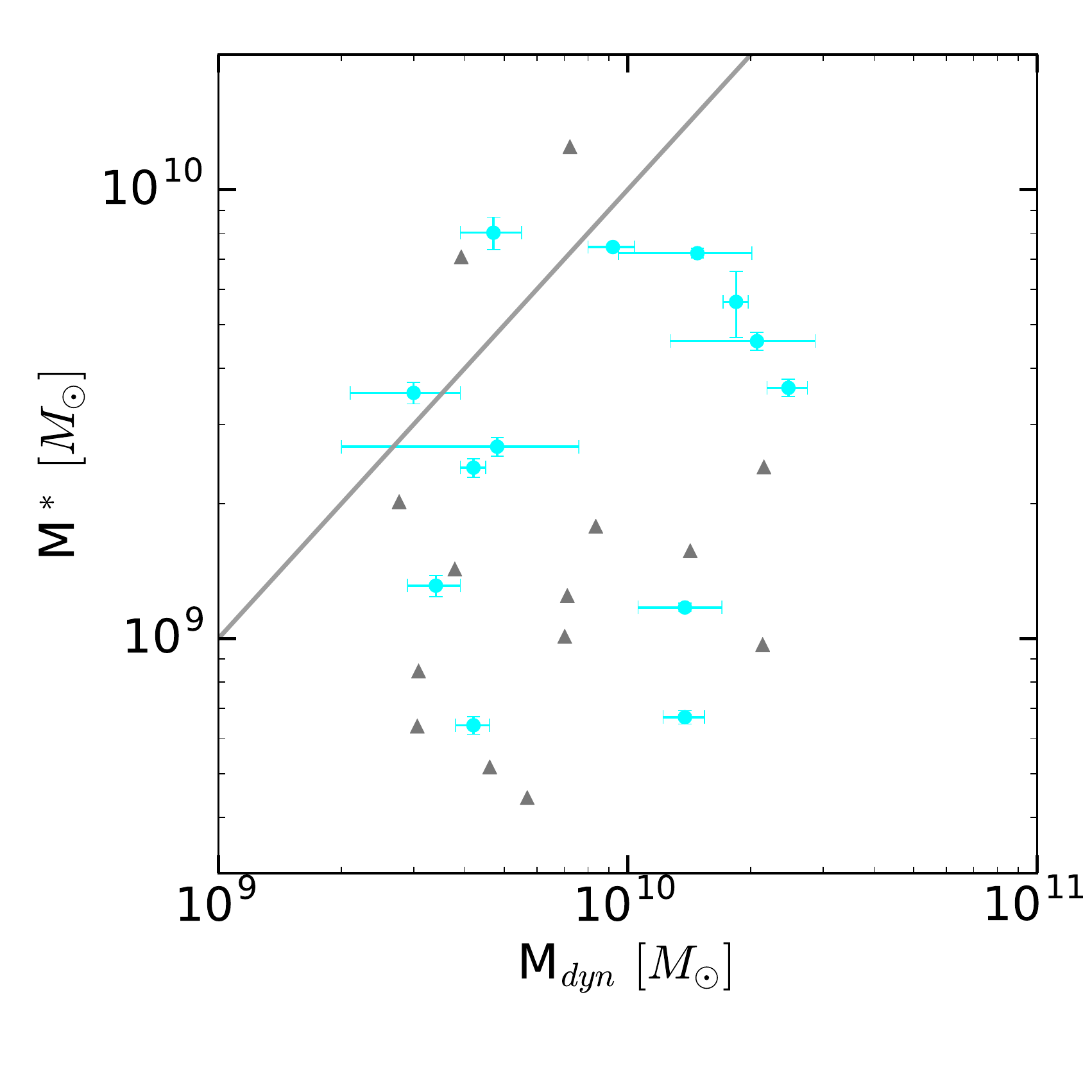}   
\caption{Comparison of best-fit BC03 stellar masses (M$_{\ast}$) and dynamical masses ($\Mdyn$) of cluster  LCBGs (teal circles) to the field sample (grey triangles). The black dashed line indicates the one-to-one line i.e. equal mass. Error bars correspond to average 1$\upsigma$ errors.
} 
\label{fig:DynStellar} 
\end{figure}

In Fig.~\ref{fig:DynStellar}, we plot stellar masses (derived from fitting SED models to the photometry) versus dynamical masses determined from the widths of emission lines for both our cluster LCBG sample and the \cite{2003ApJ...586L..45G} field LCBG comparison sample. Stellar mass measurements can serve as an estimator of the total baryonic mass and offer a good ``snapshot'' of the star formation history of these compact star-forming galaxies. In contrast, dynamical mass estimates serve as an excellent tracer of the underlying dark matter halo.  The ratio of baryonic-to-dynamical mass derived from these measurements is a key indicator of the presence of dark matter.  

We find that the median $\Mdyn/\Mast = 2.6$ with a median absolute deviation of 1.7 for cluster LCBGs as compared to $\Mdyn/\Mast=4.8$ with a median absolute deviation of 4.0 in the field sample.  This suggests  that the field LCBG population is more highly dominated by dark matter than the cluster sample.   The distribution of the ratios is presented in the right-hand panel of Figure \ref{fig:Mstar-cluster-field}.  While the K-S test does not rule out the null hypothesis, a random sampling of the field values only reproduces the median cluster value in 3.3\% of the samples.

We find that the baryonic mass exceeds the dynamical mass (i.e., $\Mast>\Mdyn$) for several targets. In our cluster sample, for which we have derived measurements errors, only one galaxy has a baryonic mass greater than the dynamical mass by more than 1$\sigma$ (standard error). We interpret this as indicating existence of relatively modest systematic errors in our mass estimates.  On the other hand, galaxies strongly dominated by dark matter will have a  typical upper limit on the ratio of stellar to dynamical mass of  $\Mdyn/\Mast<3.3$  (i.e., $\Mdyn\geq3.3~\Mast$) \citep{2014MNRAS.440.1634P}. Most of the field sample is above this limit while only a few cluster sources show so a high ratio.  This further supports the conclusion of difference in the dynamical to stellar mass ratios of the two populations.

\subsubsection{Specific Star Formation Rate}  

\begin{figure} 
\includegraphics[width=0.45\textwidth]{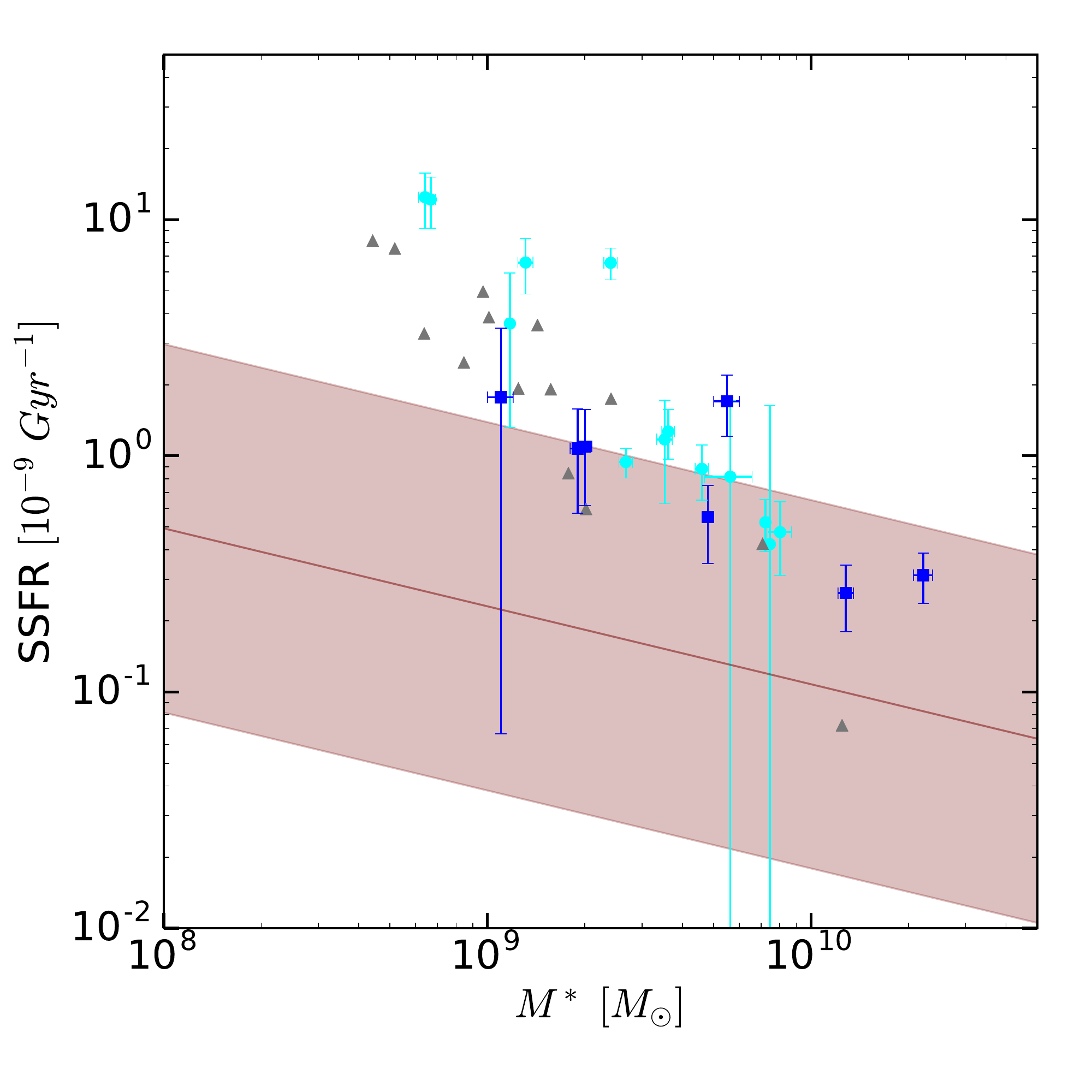}
\caption{The specific SFR as a function of  stellar mass for cluster LCBGs (teal circles), cluster blue star forming galaxies (blue squares), and field LCBGs (grey triangles).  Error bars represent $1\sigma$ estimates of random errors.   For comparison, the pink shaded region depicts the expected sSFR-stellar mass relationship \citep{2007ApJ...660L..43N}.  }

\label{fig:Mstar_sSFR} 
\end{figure}

Figure~\ref{fig:Mstar_sSFR} shows the specific star formation rate (sSFR, defined as the star formation rate per unit mass) versus stellar mass. The star formation rate is measured from the [\Oii]$\lambda3727$ emission lines following \citep{1997ApJ...489..559G,  2016ApJ...817...87C}.  The conversion from [\Oii]$\lambda3727$ equivalent width, EW$_{3727}$ to star formation rate is given by: 
$$SFR(M \ yr^{-1})=2.5 \times 10^{-0.4(M_B-M_{B \odot})} EW_{3727}$$
where M$_B$ is the absolute B-band magnitude. This star formation rate is a factor of ~3 less than the H$\alpha$ conversion from \cite{1992ApJ...388..310K}.  Star formation rates from other literature sources have been converted to this scale.  We note a decline in the sSFR as the stellar mass increases for the cluster and field LCBGs. 

For comparison, we plot the star formation vs. stellar mass relationship from \cite{2007ApJ...660L..43N}, which covers a similar redshift range of $0.2<z<0.7$.    We observe that at this epoch, both the cluster and the field LCBG populations were forming stars more vigorously than the typically galaxy at the same stellar mass.  Furthermore, the figure indicates that  the sSFR of LCBGs was decreasing more rapidly than the galaxy main sequence with increasing stellar mass.    
   
\begin{figure}
\centering
\includegraphics[width=0.45\textwidth]{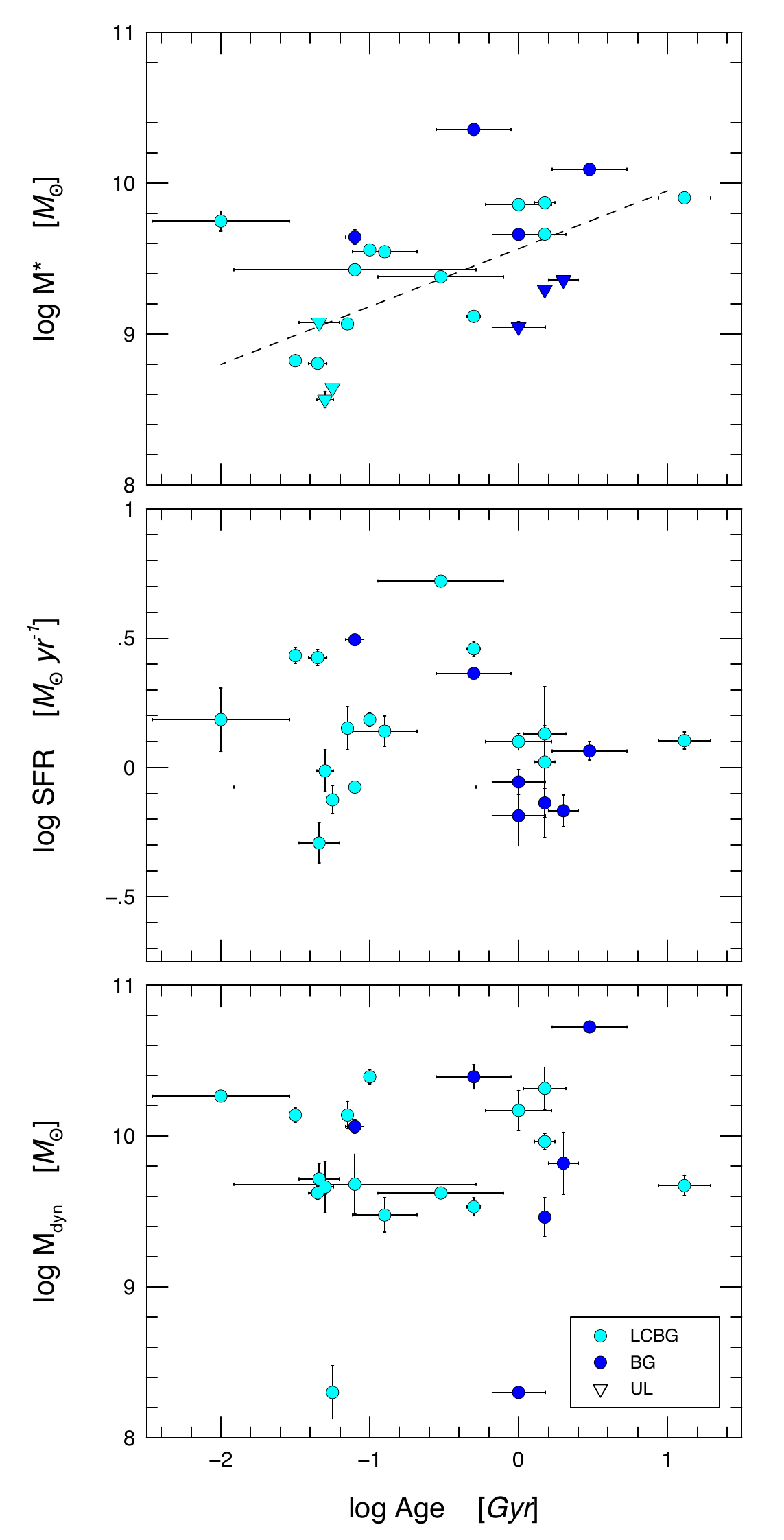}
\caption{Trends with light-weighted stellar age for best-fit BC03 stellar mass, star-formation rate, and dynamical mass (top to bottom panels) for cluster LCBGs and BGs. Uncertainties in stellar age are taken as half the difference between best-fit BC03 and CN09 values. Only stellar mass shows a significant correlation with stellar age, which drives a similar correlation in specific-SFR but only when normalized by stellar mass. The dashed line in the top panel indicates the change in mass with stellar population age for a source of constant luminosity, i.e., the line represents the relative trend in rest-frame 1.4$\mu$m M/L with age for the SPS models fit to the data.} 
\label{fig:age_trends} 
\end{figure}

\subsection{Trends with stellar age}

We find that  the ratio of stellar to dynamical mass appears to  have a significant correlation with the inferred age  of the stellar population. As seen in Figure~\ref{fig:age_trends}, this correlation is actually between the stellar mass and age, as one might expect. The correlation is driven by systematic increase in the model mass-to-light ratio for older stellar populations (at constant luminosity). The scatter about the correlation between stellar mass and age reflects the range of luminosities in the sample. Surprisingly, the light-weighted mean stellar age does not appear to correlate with the current star-formation rate (SFR), as inferred by the [\Oii] luminosity. This implies that galaxies with older mean ages for their stellar populations must also have, on average, more stellar mass from previous generations of star-formation.

\section{Discussion} 
\label{sec:Discussion}

Several studies suggest that LCBGs may be the progenitors of dwarf elliptical galaxies \citep{1994ApJ...430..121C,1997ApJ...478L..49K,1998ApJ...495..139M,2005ApJ...619..134T,2016ApJ...817...87C}.  The assumed mechanism for their formation is that upon entering a cluster for the first time, infalling galaxies experience a burst of star formation that is suddenly quenched. In the process, these objects  lose significant mass through stripping.  These galaxies then passively  evolve into the large population of low-mass spheroidal galaxies dominating clusters today. 

In this paper, our main findings indicate similar stellar mass distributions between cluster and field LCBGs, in accordance with previous findings that various properties of LCBGs (size, luminosity, dynamical mass, SFR, and metallicity) in  cluster and field populations are indistinguishable \citep{2016ApJ...817...87C}.  

We also find a lower dynamical-to-stellar masses ratio for cluster LCBGs as compared to the field, although a wide range of values exist for individual objects.  Nonetheless, the values found for cluster ($2.6$) and field ($4.8$) LCBGs  -- both relative and absolute --  are consistent with the individual values found for cluster ($2.2\pm0.5$) and field ($5.1\pm0.6$) dwarf ellipticals \citep{2015MNRAS.453.3635P}.  They are also similar to the cluster values ($2.5\pm0.25$) found in the interior of Virgo by \cite{2014MNRAS.439..284R}.  Furthermore, the relationship for LCBGs shows the same pattern: field dE have a higher dynamical-to-stellar mass ratio than cluster dE \citep{2015MNRAS.453.3635P}.  This would imply little evolution over time in the dynamical to stellar mass ratio for dwarf galaxies in clusters, which may be in some conflict with predictions from simulations \citep{2016MNRAS.455.2323M}.

The LCBG population features very low mass galaxies forming stars at extreme levels for their stellar mass.  Much of their observed light is dominated by this burst in star formation, with some of the sources being undetected in the $K$-band despite having high luminosities.  In addition, some LCBGs have also shown evidence for having a large amount of obscured star formation \citep{2016ApJ...817...87C}.  Along with strong correlations between inferred age and stellar mass seen when fitting stellar populations models, a more detailed study of the spectral energy distributions for these galaxies is needed to better understand their recent and past star formation.   Obtaining accurately flux calibrated spectra for these sources will also enable better comparison with stellar population models.

Our overall sample and much of the statistical results presented here are limited to a small sample from a single survey.  Extending this type of analysis to more clusters and a range of environments is necessary to confirm the results that we have found here and to further explore how the stellar mass and star formation history of these galaxies change with environment and time.

\section{Conclusion} \label{sec:Conclusion}

In this paper, we have measured the stellar masses of cluster LCBGs and cluster blue star-forming galaxies at intermediate redshift (z $\approx$ 0.54). To do this, we performed broadband SED-fitting by comparing a grid of template galaxy SED models to the observed broadband SEDs.   We generated SED models using the public stellar population synthesis codes of BC03 (GALAXEV) and CN09 (FSPS), both of which span a wide range of inputs for age, star formation history, extinction, and metallicity.  

We estimated the stellar masses based on the SED best-fitting results and found consistent agreement in stellar masses when using the two SED models. We also inferred other physical properties of each galaxy, and have recorded these results derived through the SED-fitting process that include stellar masses, age, star formation history, extinction, and metallicity. The dynamical masses of the cluster LCBGs were calculated from their velocity dispersion measurements.

In particular, we have compared the stellar masses and dynamical masses of cluster LCBGs to their field counterparts.  While the two samples have similar distributions, we have found a lower dynamical-to-stellar mass ratio for cluster LCBGs compared to the field, which is consistent with the distributions seen for low-redshift dwarf elliptical galaxies.   If LCBGs are the progenitors of low redshift dE galaxies, this would imply that the ratio of dynamical-to-stellar-mass is set when the galaxy is falling into the cluster for the first time.

\section*{Acknowledgments} 
We do thank the anonymous referee for his/her valuable and constructive comments that improved the quality of the paper. 
S.M.R. wishes to thank the SAAO/UWC and UKZN for their supports towards this research project. The financial assistance of the South African SKA Project (SKA SA) towards this research is hereby acknowledged. Opinions expressed and conclusions arrived at are those of the authors and are not necessarily to be attributed to the SKA SA (www.ska.ac.za).  S.M.C. acknowledges the South African Astronomical Observatory and the National Research Foundation of South Africa for support during this project.  M.A.B. acknowledges support from NSF AST-1517006.
 
This research made use of Astropy, a community-developed core Python package for Astronomy \citep{2013A&A...558A..33A}, scipy \citep{scipy}, and matplotlib \citep{Hunter:2007} .  We are grateful to the WIYN Observatory and the W. M. Keck Observatory for the spectroscopic observations. The authors wish to recognize and acknowledge the very significant cultural role and reverence that the summit of Mauna Kea has always had within the indigenous Hawaiian community.  We are fortunate to have the opportunity to conduct observations from this mountain. 
%



\bibliographystyle{mnras}  
\bibliography{smass_paper}



\appendix
\section{SED-fitting results} \label{sec:app}
In this section, we present the SED best-fit figures and a table that display the best-fit results estimated from the SED models which include metallicity, age, extinction, SFH, and stellar masses. We note that in all figures CL is the same as CZ in the text which refers to \cite{2000ApJ...533..682C} extinction law. \\

\begin{center}
\begin{table*}
\centering
\caption{Best-fit physical parameters and $\chisq$ for BC03 and CN09 SED models.} 
\label{tab:params003} 
{\scriptsize
\begin{tabular}{c  l l l l l l l l l l l l l l l l l  } \hline \hline 
 & \multicolumn{6}{c}{BC03} & & \multicolumn{6}{c}{CN09} & & \\ \cline{2-7} \cline{9-14}
 \multicolumn{16}{c}{}\\ [-0.05in]
Obj  & Z         & Age  & E(B-V) & SFH   & SFH  & M$_{*}$         && Z         & Age & E(B-V)& SFH  & SFH & M$_{*}$ & $\chisq$ & $\chi^2$ \\ 
(ID) &(Z$_\odot$)& (Gyr)&        & (type)& (Gyr)&($10^{9}\ \Msun$)&&(Z$_\odot$)&(Gyr)&      & (type)&(Gyr)&($10^{9}\ \Msun$)& BC03 & CN09 \\ 
        (1)    & (2) & (3) & (4)   & (5) & (6) & (7) & & (8) & (9) & (10)  & (11)  & (12) & (13) & (14)   & (15)   \\ \hline
\multicolumn{16}{c}{Cluster LCBGs}\\ \hline 
1081 & 1.000 &  0.0708 &  0.50---CZ & SSP & ...    & 1.173    & & 1.000  &  0.075 & 0.50---CZ & B & 0.500 & 1.004  & 7.32 & 7.52 \\
1093 & 0.400 &  0.0100 &  1.00---CZ & E   & 0.005 & 5.625    & & 0.100  &  0.048 & 1.00---CZ & E & 0.500 & 9.300  & 5.75 & 2.47 \\
1252 & 1.000 &  1.0000 &  0.25-MW   & E   & 0.010 & 7.221    & & 1.000  &  2.334 & 0.25-MW   & C & 1.000 & 9.382  & 7.62 & 5.24 \\
1348 & 0.400 &  0.0501 &  0.25---CZ & SSP & ...    & $<$0.369 & & 1.000  &  0.064 & 0.25---CZ & C & 0.500 &$<$0.399& 1.04 & 0.93 \\
1954 & 0.005 & 13.000  &  0.10-MW   & C   & 0.015 & 8.020    & & 1.000  &  0.101 & 0.50---CZ & C & 0.500 & 1.841  & 34.61 & 27.72 \\ 
1968 & 1.000 &  0.0794 &  0.25---CZ & SSP & ...    & 2.676    & & 0.400  &  0.953 & 0.10-MW   & E & 1.000 & 4.367  & 8.52 & 7.72 \\
2084 & 2.500 &  1.5000 &  0.10-MW   & E   &0.010  & 4.598    & & 0.025  &  0.335 & 1.00---CZ & E & 2.000 & 5.380  & 28.16 & 25.12 \\
2224 & 1.000 &  0.1000 &  0.50---CZ & SSP & ...    & 3.619    & & 0.025  &  0.087 & 1.00---CZ & B & 0.500 & 8.175  & 46.29 & 46.49 \\
2300 & 2.500 &  0.0447 &  0.10---CZ & SSP & ...    & 0.641    & & 1.000  &  0.031 & 0.25---CZ &SSP& ...    & 0.625  & 50.48 & 87.34 \\
2932 & 2.500 &  0.0562 &  0.10---CZ & SSP & ...    & $<$0.443 & & 0.400  &  0.064 & 0.10-MW   & C & 0.500 &$<$0.386& 15.81 & 36.36 \\
664  & 1.000 &  0.1259 &  0.50---CZ & SSP & ...    & 3.527    & & 0.025  &  0.288 & 0.50---CZ & B & 0.500 & 4.831  & 24.94 & 28.96 \\
910	 & 0.200 &  0.5000 &  0.10-MW   & E   & 0.010 & 1.312    & & 1.000  &  0.609 & 0.10-MW   & E & 1.000 & 1.676  &  5.51 & 3.48 \\
947  & 0.200 &  0.0316 &  0.10-MW   & SSP & ...    & 0.668    & & 0.200  &  0.035 & 0.10-MW   &SSP& ...    & 0.747  & 10.26 & 14.57 \\  
950  & 0.200 &  0.3000 &  0.10-MW   & B   & 0.020 & 2.403    & & 0.100  & 1.284  & 0.10-MW   & C & 1.000 & 5.207  & 25.94 & 11.63 \\
2947 & 0.020 &  0.0457 &  0.50-MW   & SSP & ...    & $<$1.194 & & 0.025  & 0.079  & 0.50-MW   &SSP& ...    &$<$1.835&147.41 & 142.59 \\ 
3201 & 0.200 &  1.5000 &  0.25-MW   & C   & 0.005 & 7.452    & & 1.000  & 2.010  & 0.25-MW   & C & 1.000 & 7.126  & 85.36 & 13.39 \\ \hline
\multicolumn{16}{c}{Cluster Blue Galaxies}\\ \hline 
1118 & 0.400 &  0.5000 &  1.00---CZ & C   & 0.020 & 22.739   & & 0.025  &  1.284 & 1.00---CZ & C & 1.000 & 40.438 & 17.98 & 11.12 \\
2312 & 1.000 &  0.0794 &  0.50---CZ & SSP & ...    & 4.406    & & 0.025  &  0.056 & 1.00---CZ &SSP& ...    & 10.058 & 22.66 & 17.09 \\
629  & 1.000 &  1.0000 &  0.25-MW   & E   & 0.005 & 4.581    & & 1.000  &  2.010 & 0.25-MW   & E & 1.500 & 6.047  & 31.82 & 28.24 \\
732  & 0.400 &  3.0000 &  0.25-MW   & C   & 0.015 & 12.397   & & 0.100  &  7.705 & 0.25-MW   & C & 1.000 & 19.481 & 20.19 & 14.60 \\
814  & 2.500 &  1.0000 &  0.50---CZ & C   & 0.010 & $<$1.116 & & 0.100  &  2.010 & 0.50---CZ & C & 1.000 &$<$1.198& 1.67 & 1.23 \\
925  & 0.005 &  2.0000 &  0.10-MW   & E   & 0.020 & $<$2.292 & & 0.025  &  0.953 & 0.10-MW   & E & 0.500 &$<$1.668& 7.65 & 11.05 \\
3861 & 0.020 &  1.5000 &  0.10---CZ & E   & 0.010 & $<$1.979 & &  0.025 & 1.284  & 0.00-MW   & E & 0.500 &$<$1.836& 49.11& 58.04 \\ \hline 
\end{tabular} 
}
\begin{flushleft}
Note. -- (1) Object identification number. BC03 best fit parameters: (2) Metallicity, (3) Age, (4) Extinction values and models: CZ=Calzetti, MW= Milky Way, (5) Star-formation history types: SSP=Single Stellar Population, B=Burst, C=Constant, E=Exponential, (6) Star-formation values, (7) Stellar masses. CN09 best fit parameters: (8) Metallicity, (9) Age, (10) Extinction values and models: CZ=Calzetti, MW= Milky Way, (11) Star-formation history types: SSP=Single Stellar Population, B=Burst, C=Constant, E=Exponential, (12) Star-formation values, (13) Stellar masses. (14) BC03 chi-squared values. (15) CN09 chi-squared values. Sources detected below the K-band limit (20.2 mag) of \cite{2007ApJ...671.1503M} are marked with ($<$) in front of their stellar masses. 
\end{flushleft}
\end{table*}
\end{center}

\begin{figure*}
\centering
  \begin{tabular}{@{}cc@{}}
    \includegraphics[width=.5\textwidth]{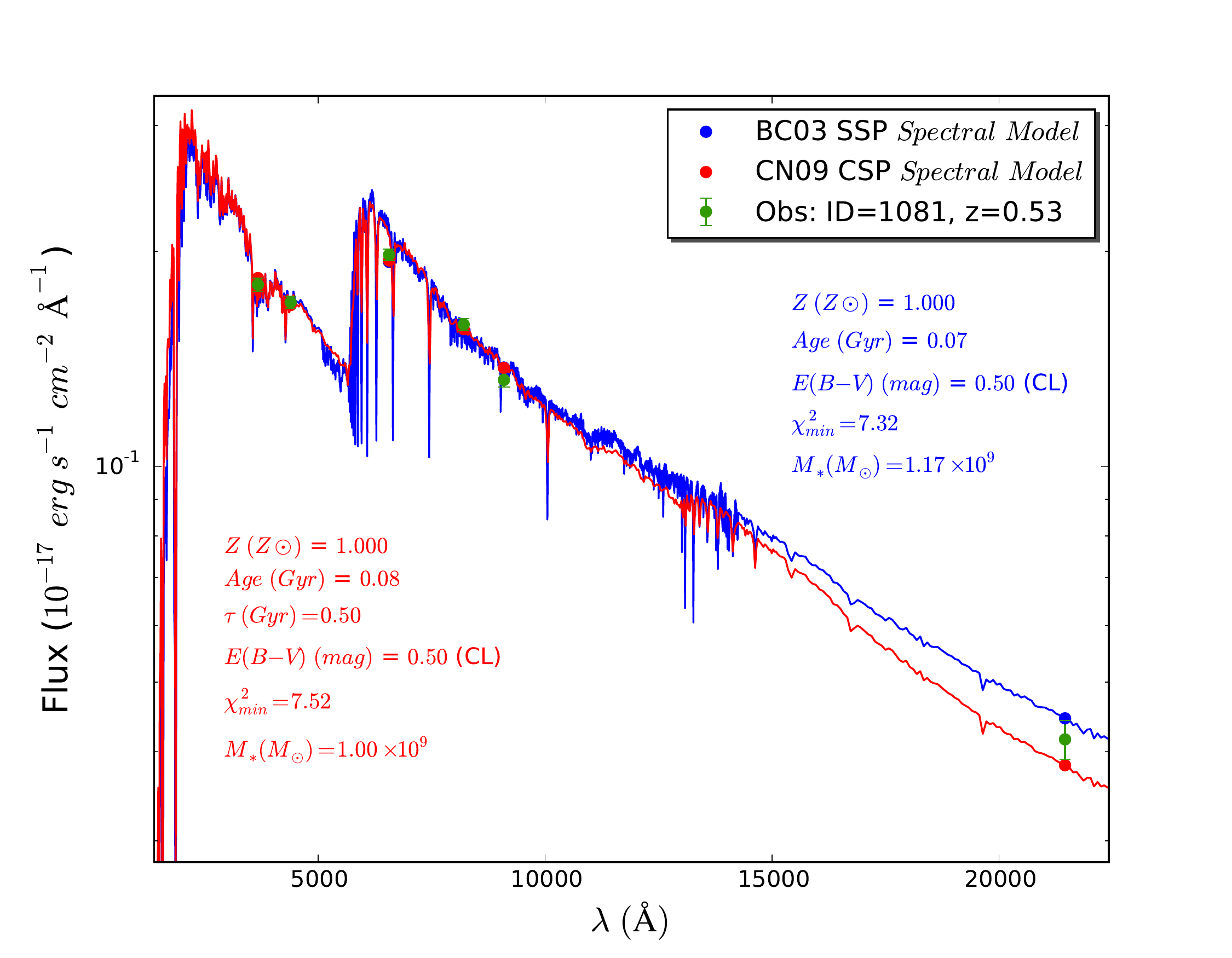} &
    \includegraphics[width=.5\textwidth]{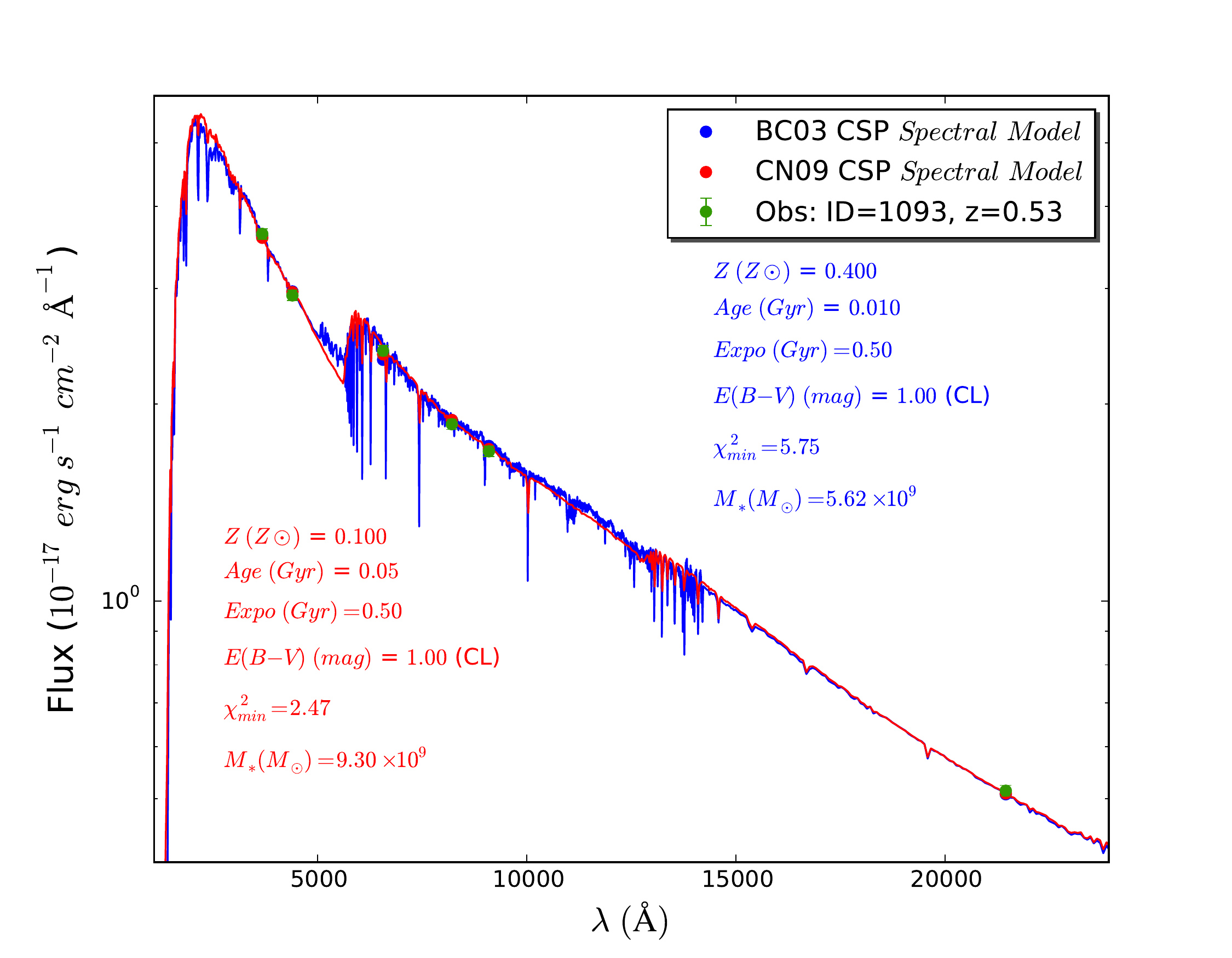}   \\
    \includegraphics[width=.5\textwidth]{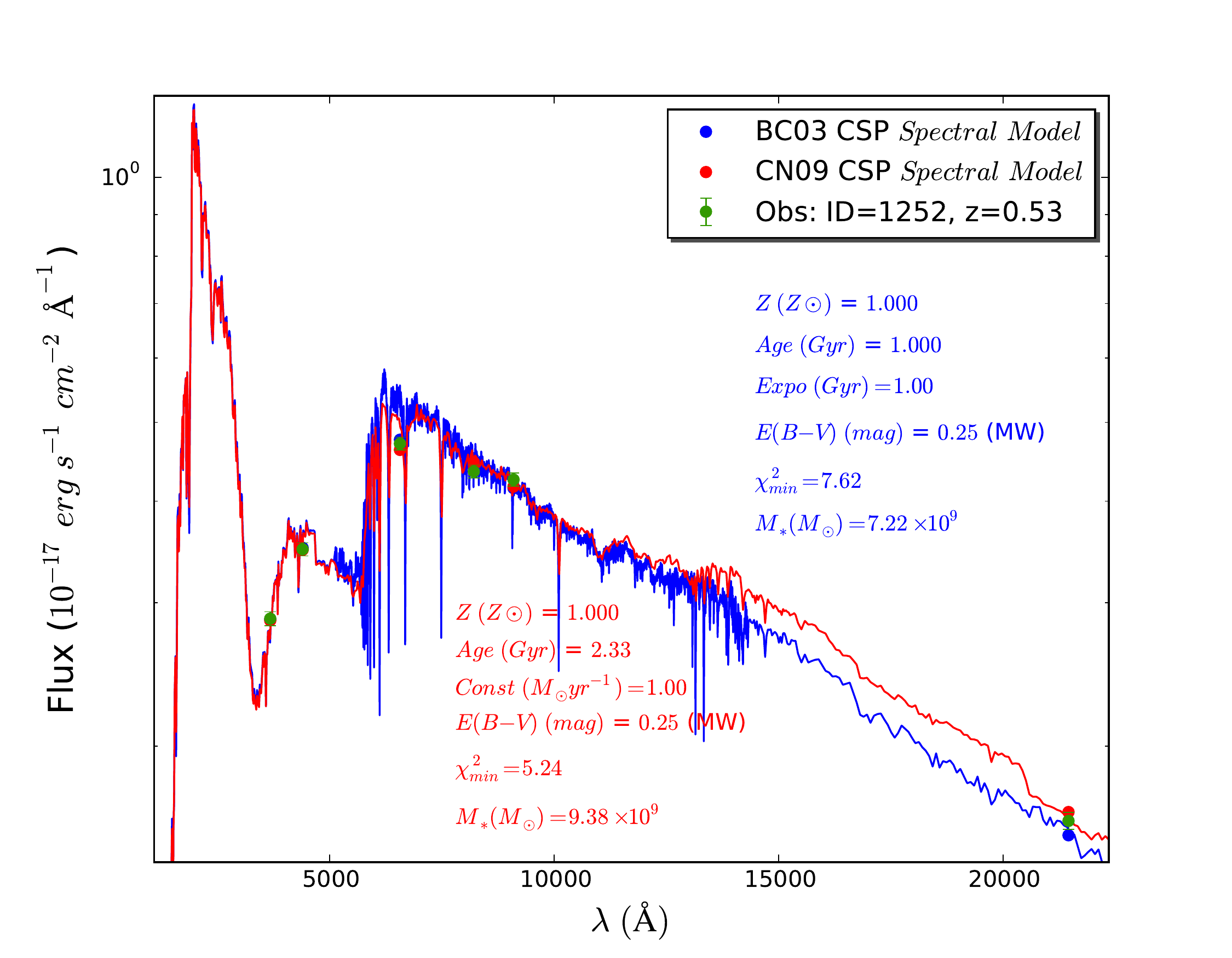}   &
    \includegraphics[width=.5\textwidth]{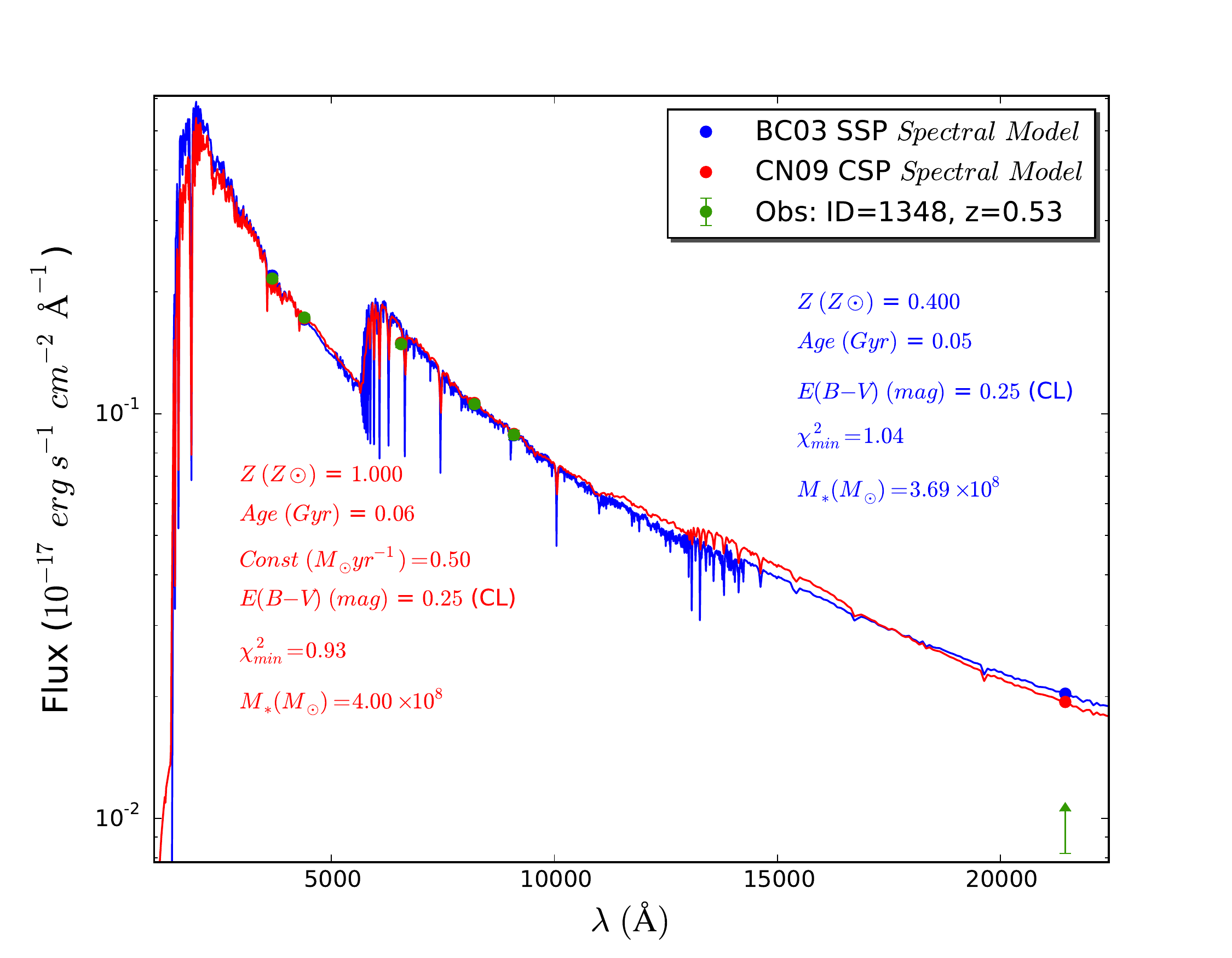}  \\
     \includegraphics[width=.5\textwidth]{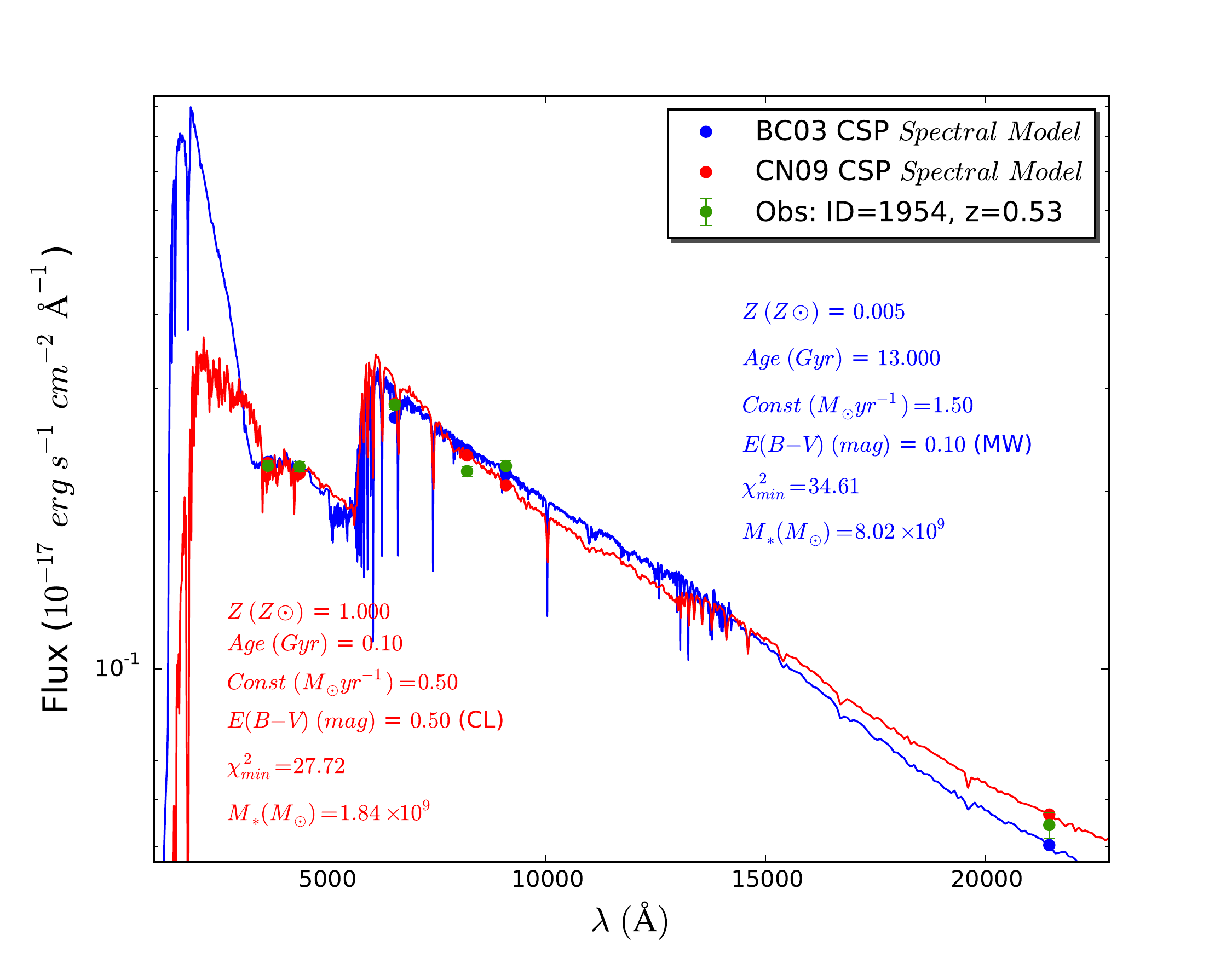} &
    \includegraphics[width=.5\textwidth]{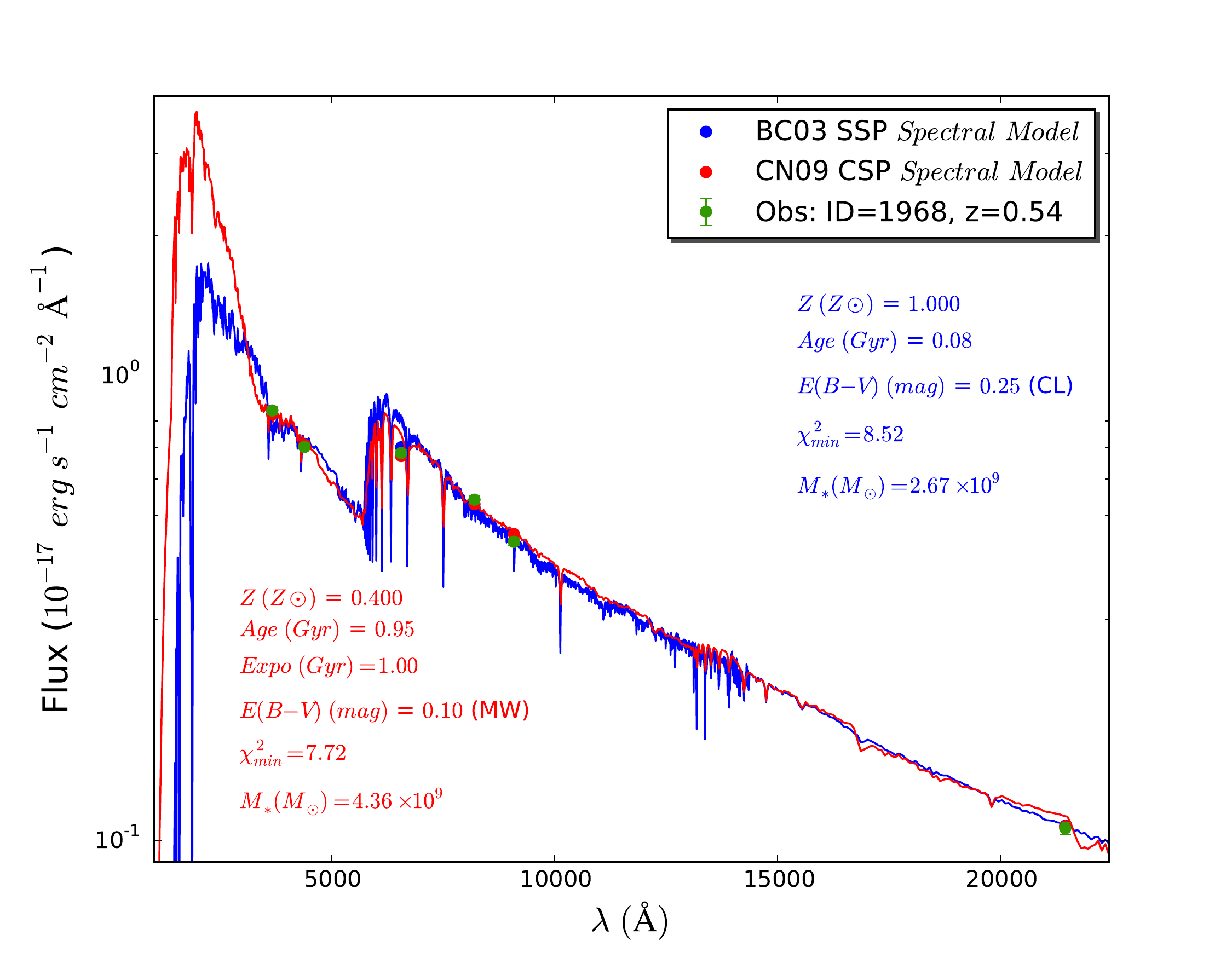}   
   \end{tabular}
   \caption{The best SED-fitting results for each sources in the sample.}
   \label{fig:A1}
  \end{figure*}

\begin{figure*}
\centering
  \begin{tabular}{@{}cc@{}}
    \includegraphics[width=.5\textwidth]{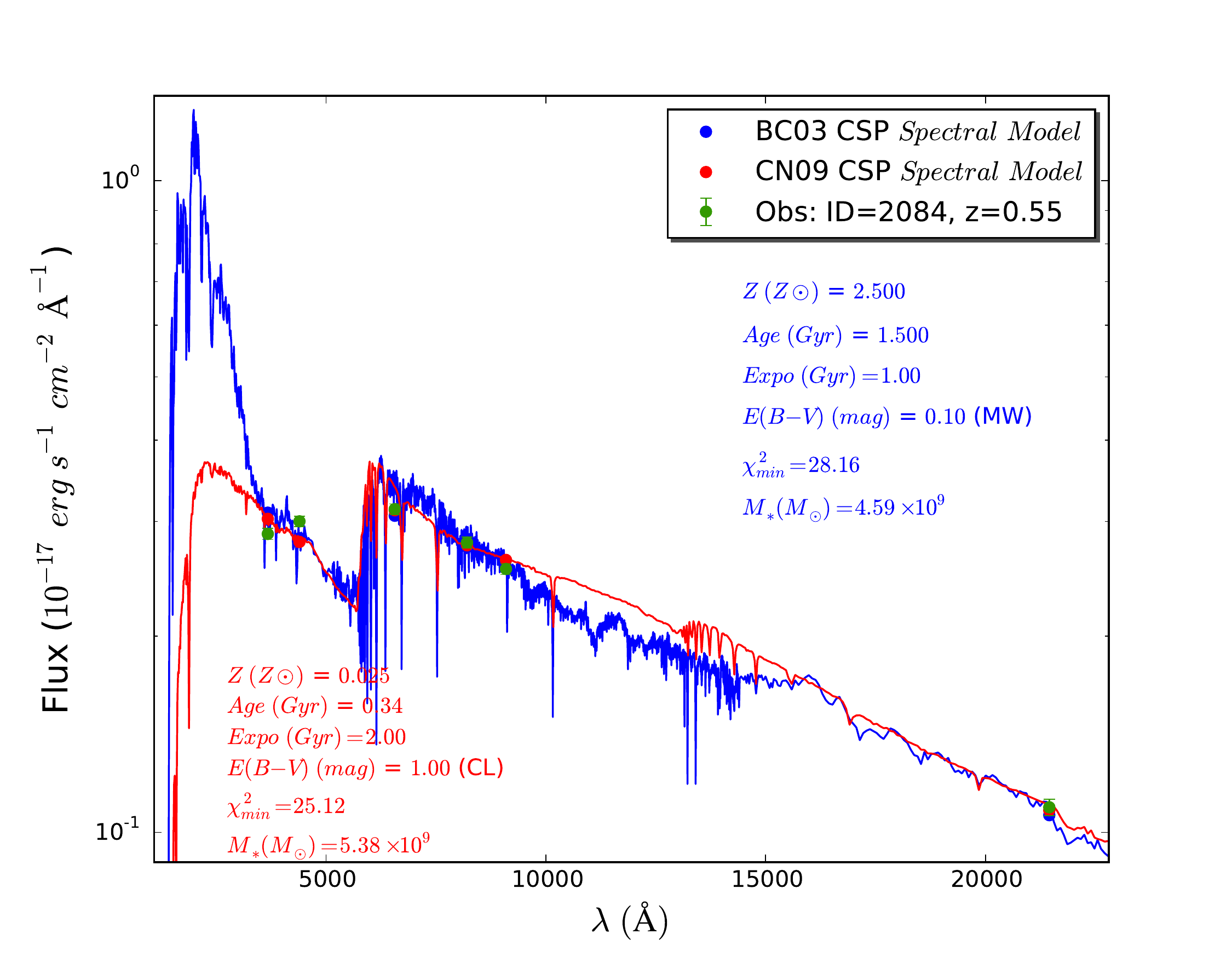} &
    \includegraphics[width=.5\textwidth]{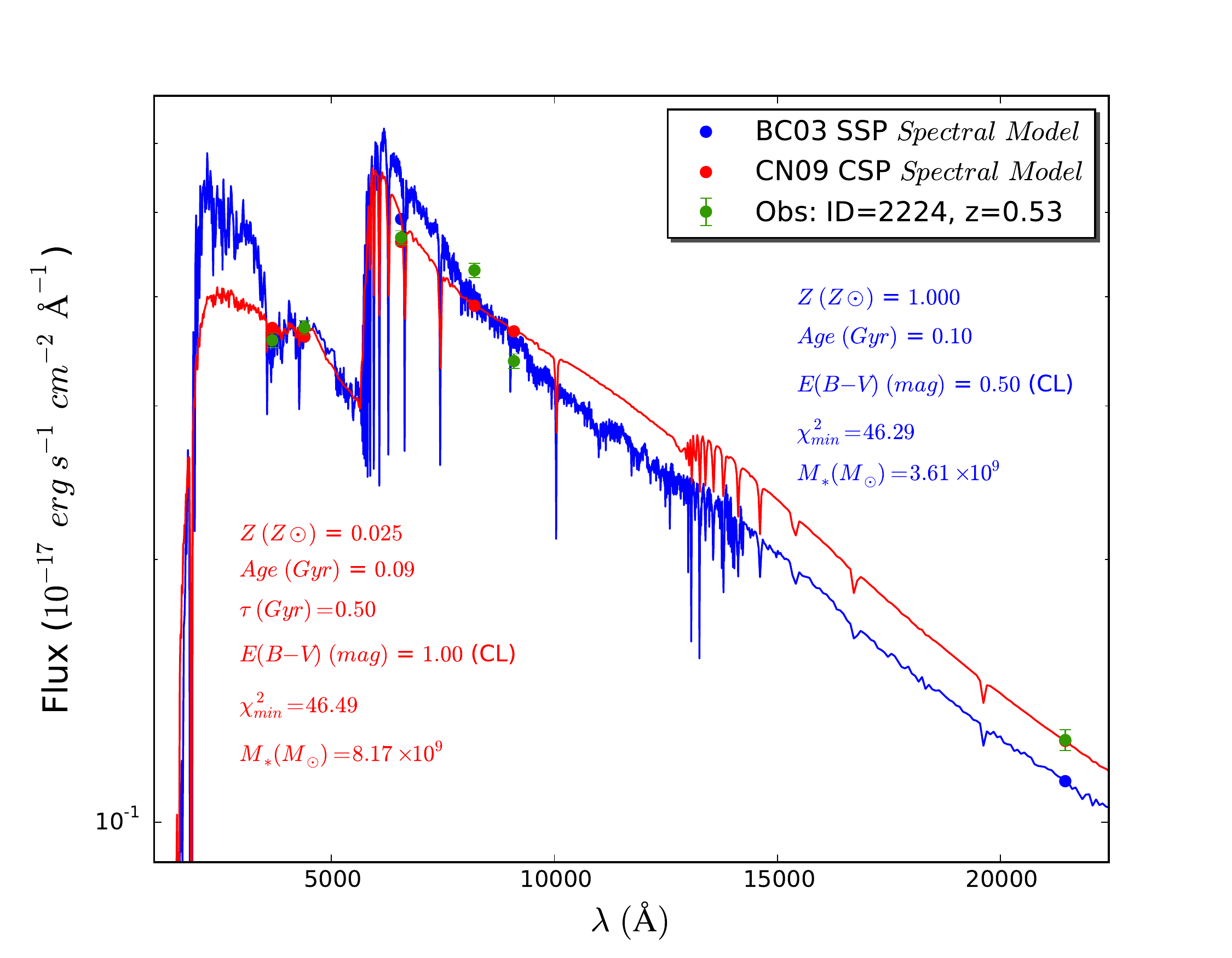}   \\
    \includegraphics[width=.5\textwidth]{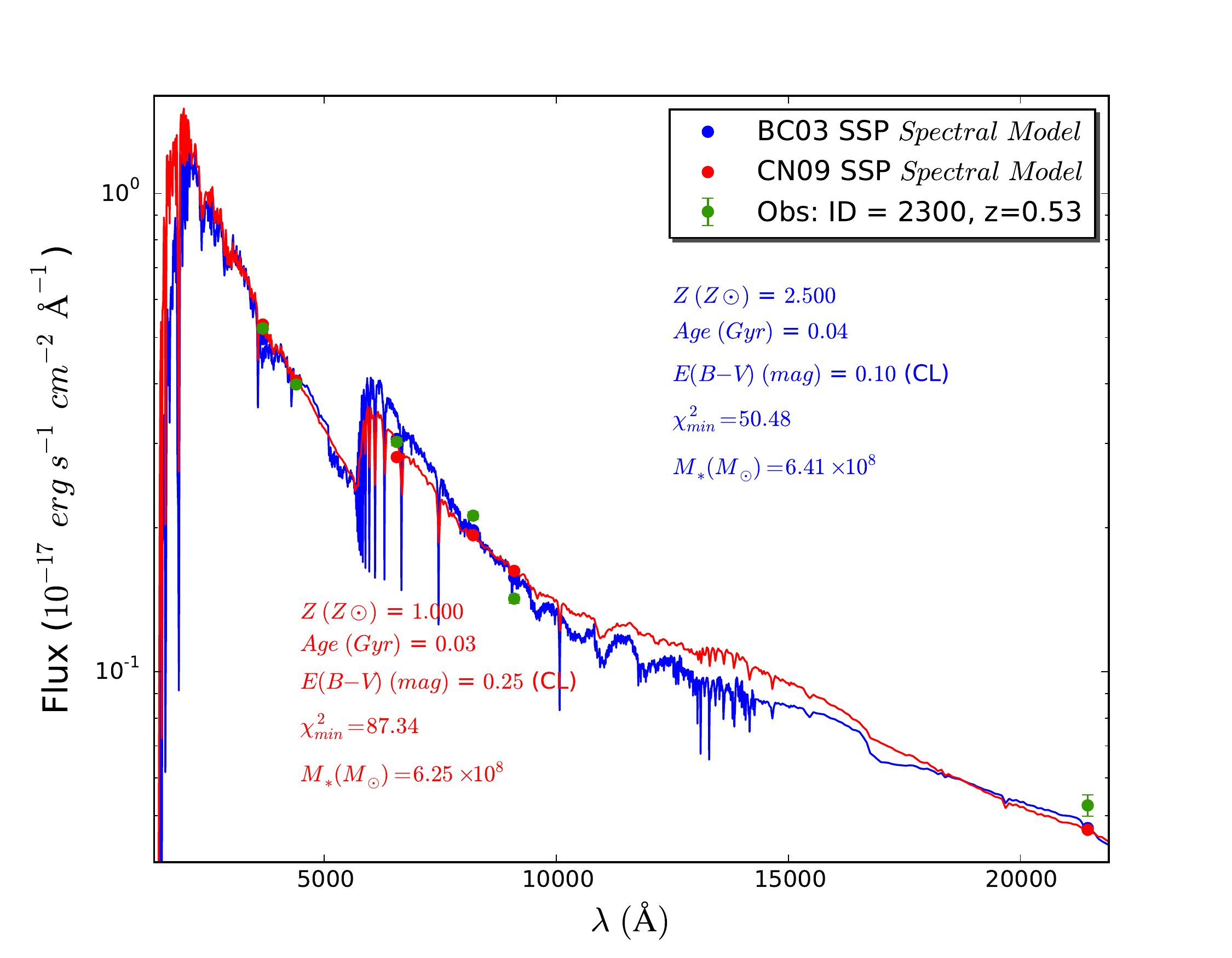}   &
    \includegraphics[width=.5\textwidth]{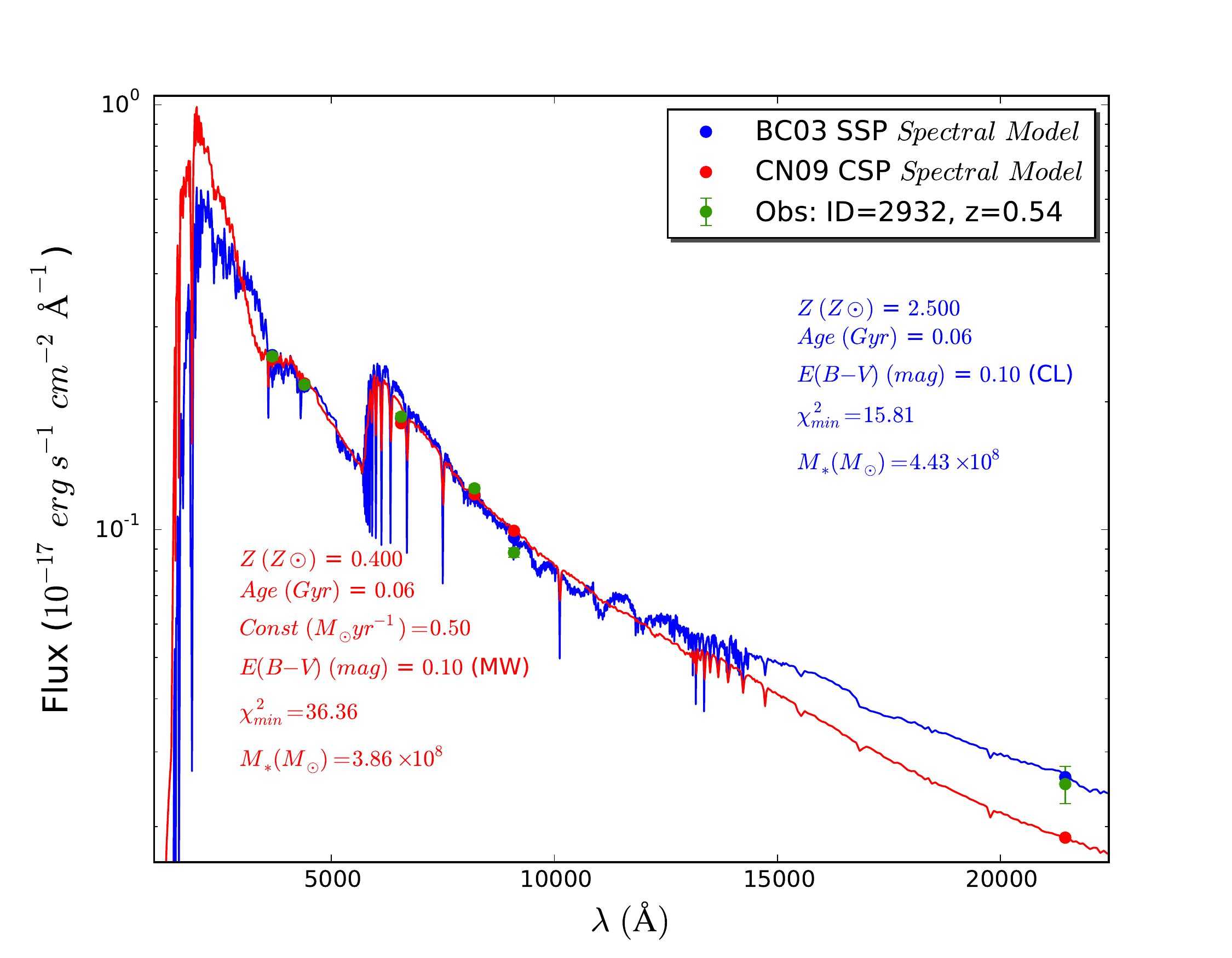}  \\
     \includegraphics[width=.5\textwidth]{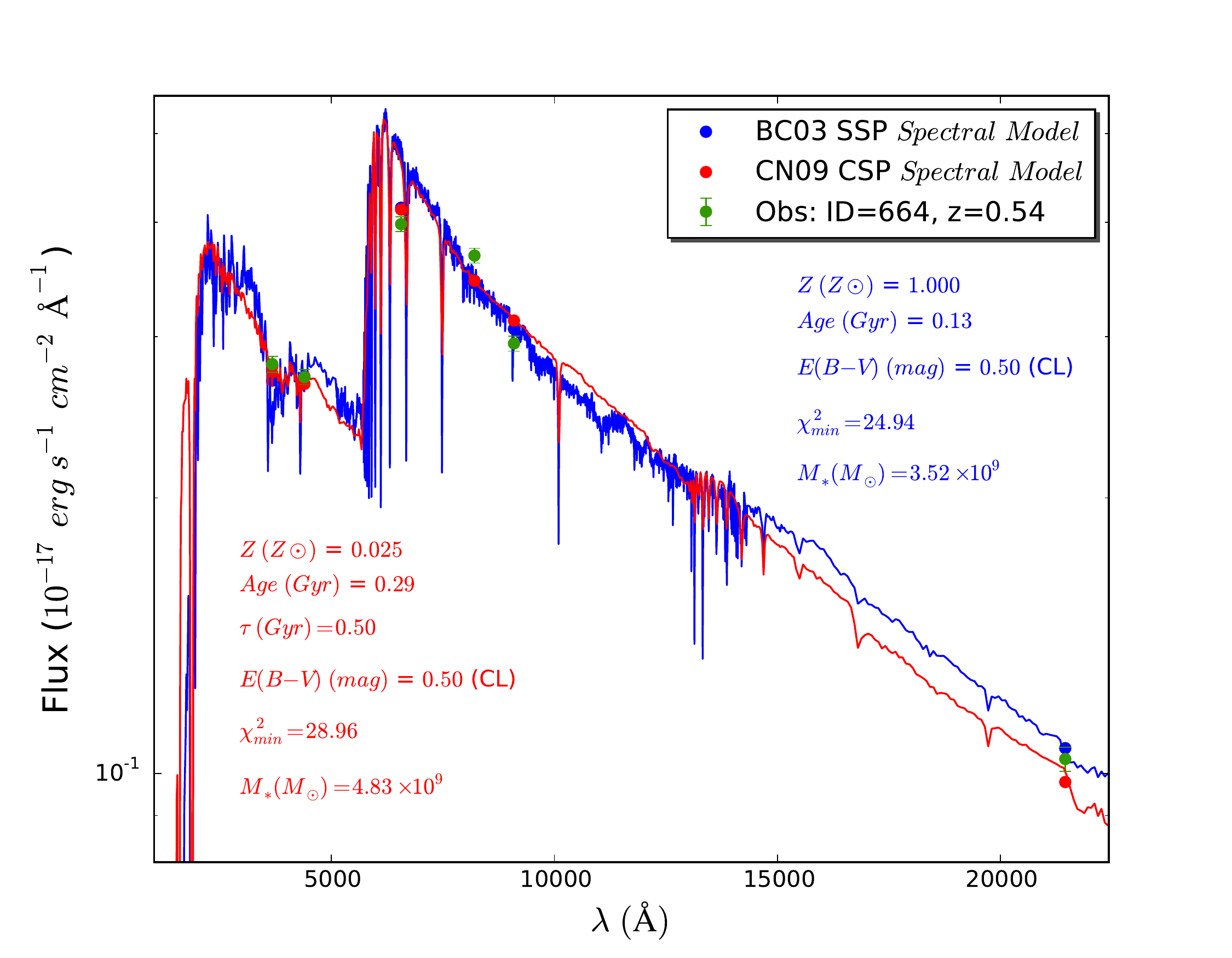} &
    \includegraphics[width=.5\textwidth]{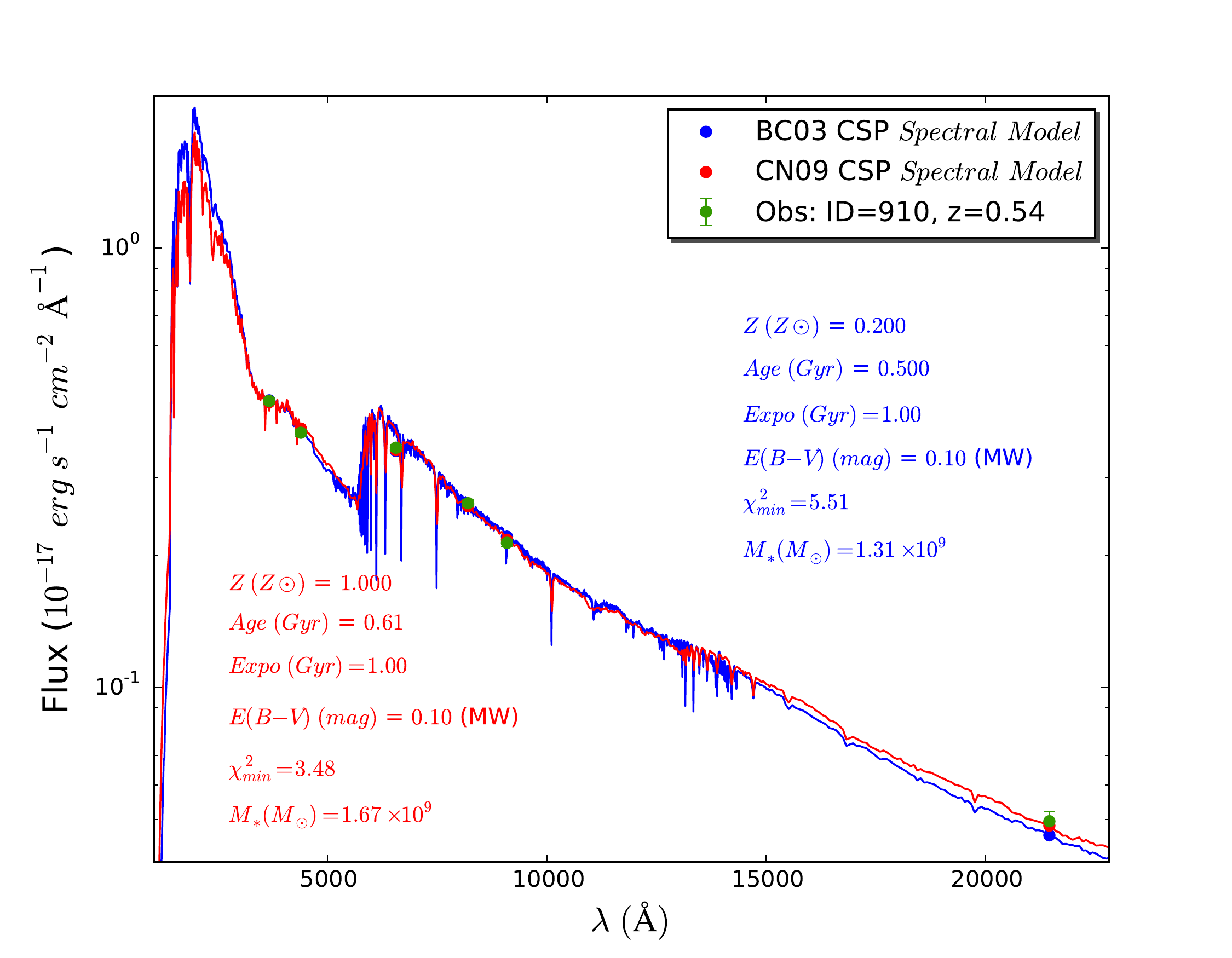}   
   \end{tabular}
   \caption{(continued)}
      \label{fig:A2}
  \end{figure*}
  
  \begin{figure*}
\centering
  \begin{tabular}{@{}cc@{}}
    \includegraphics[width=.5\textwidth]{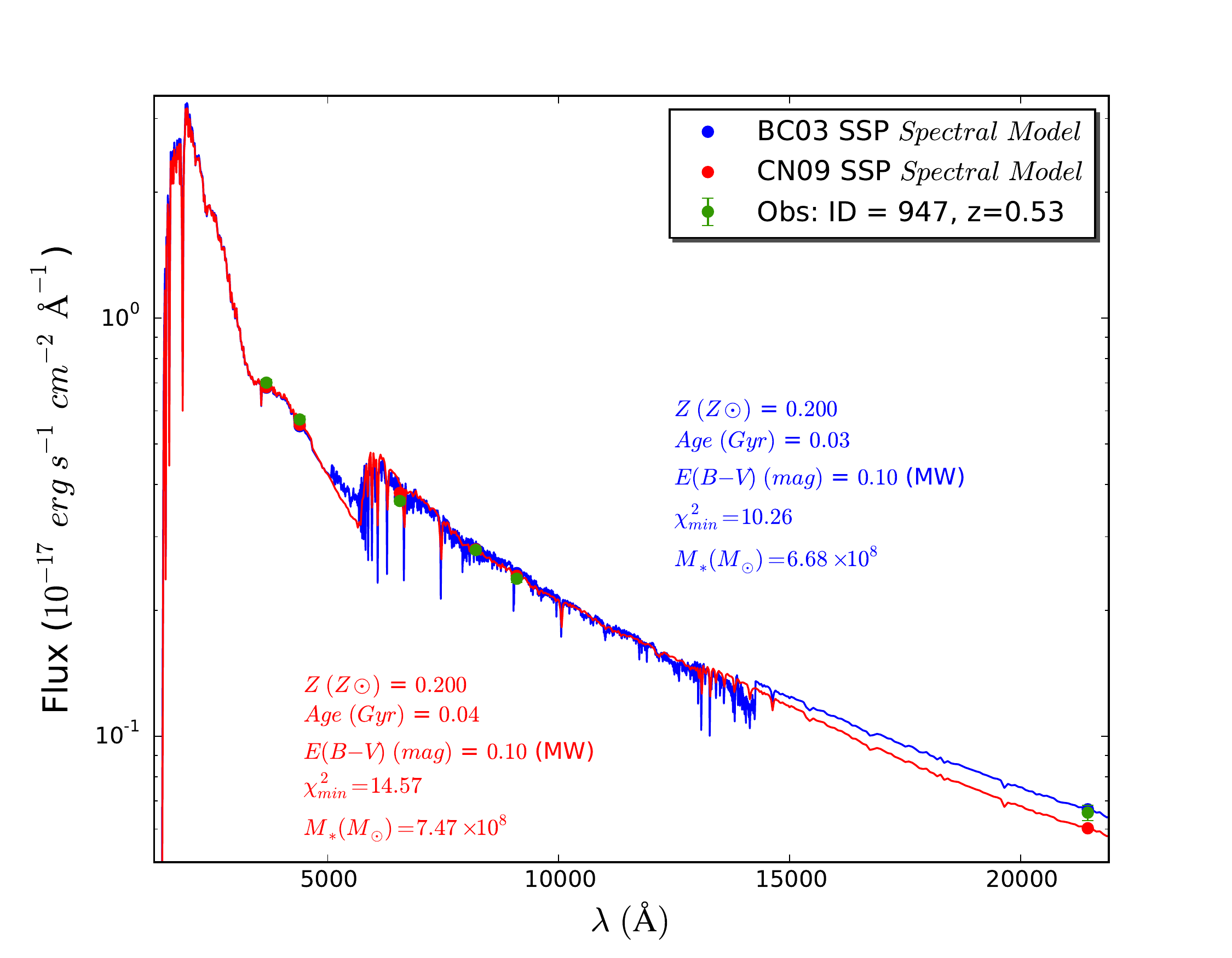} &
    \includegraphics[width=.5\textwidth]{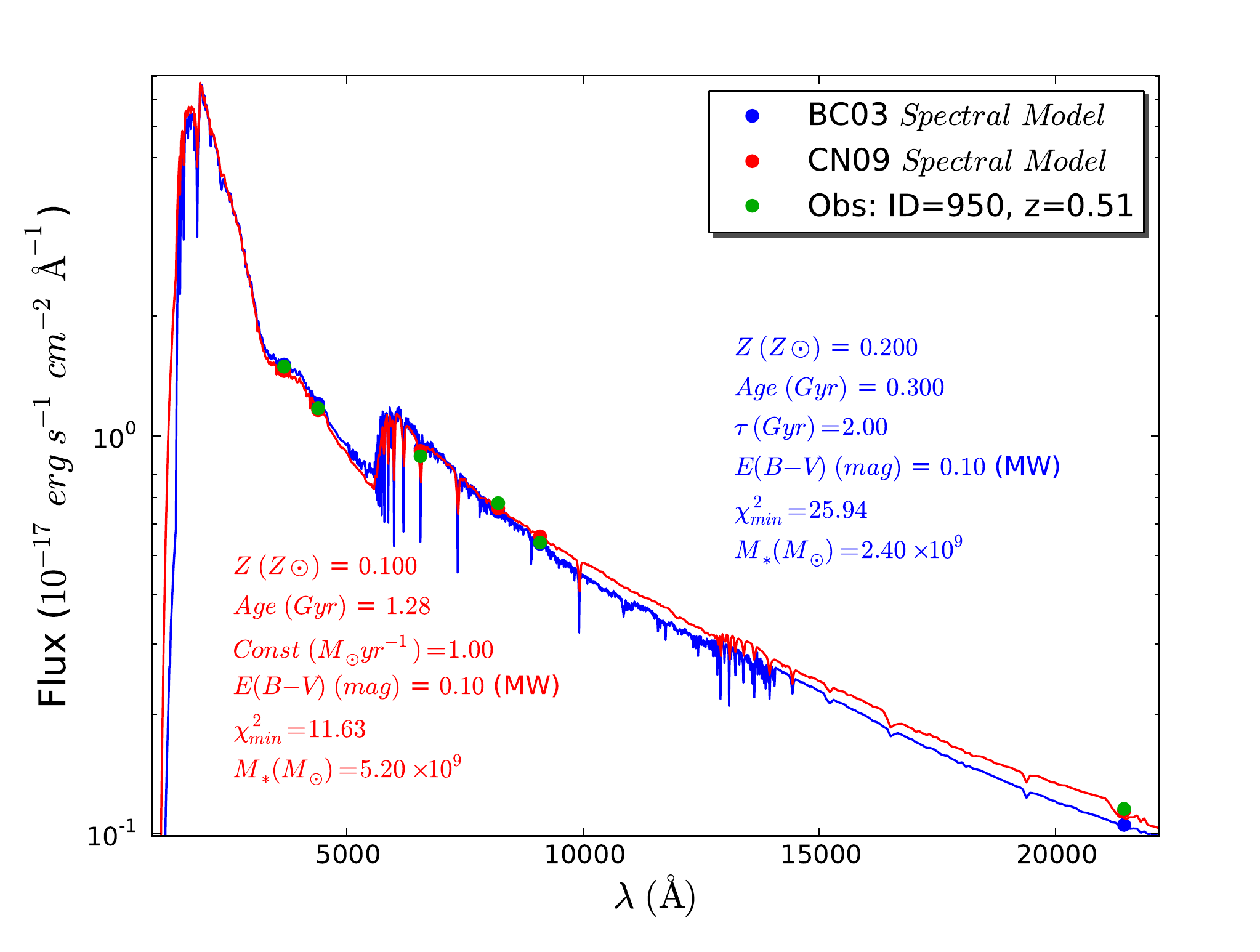}   \\
    \includegraphics[width=.5\textwidth]{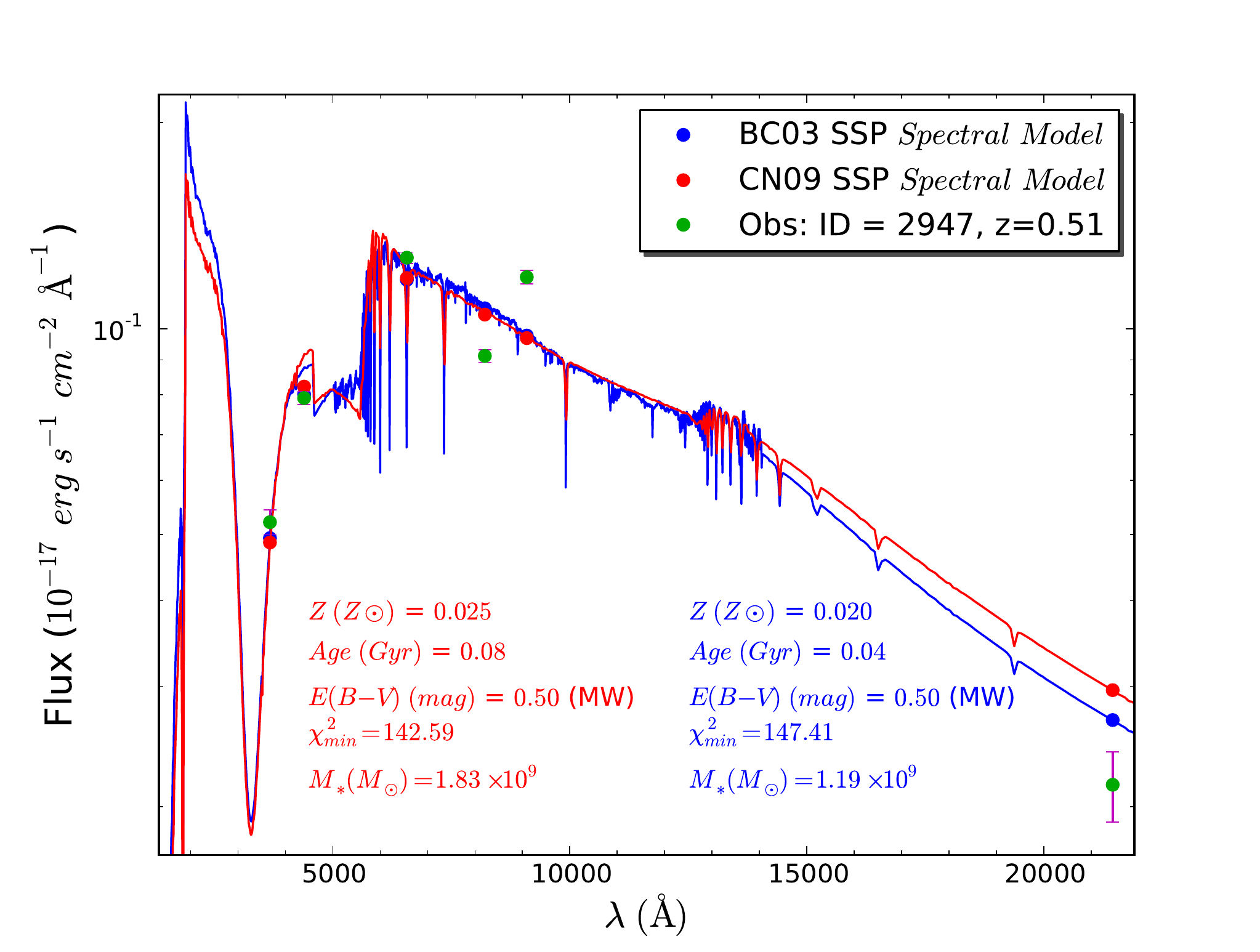}   &
    \includegraphics[width=.5\textwidth]{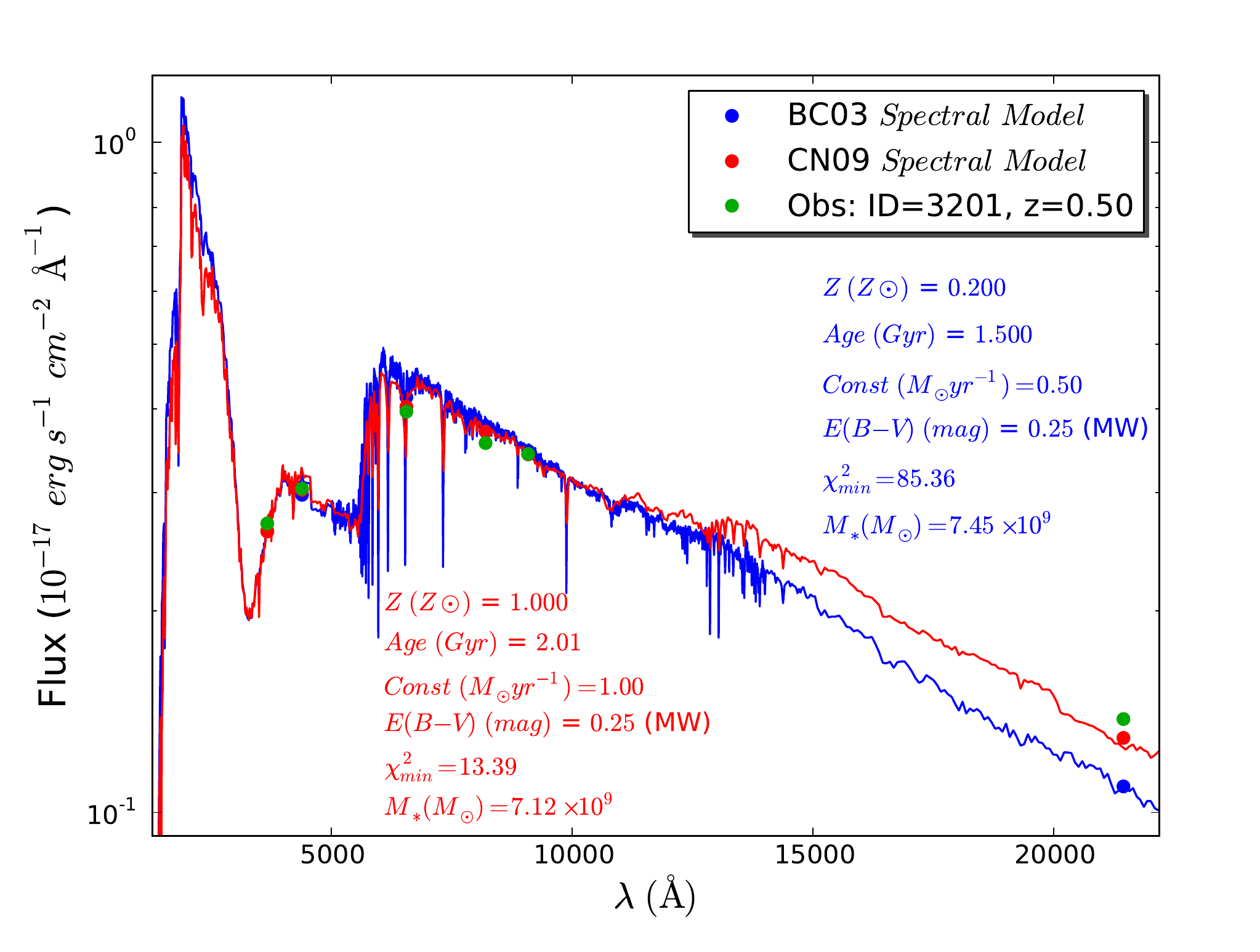}  \\
     \includegraphics[width=.5\textwidth]{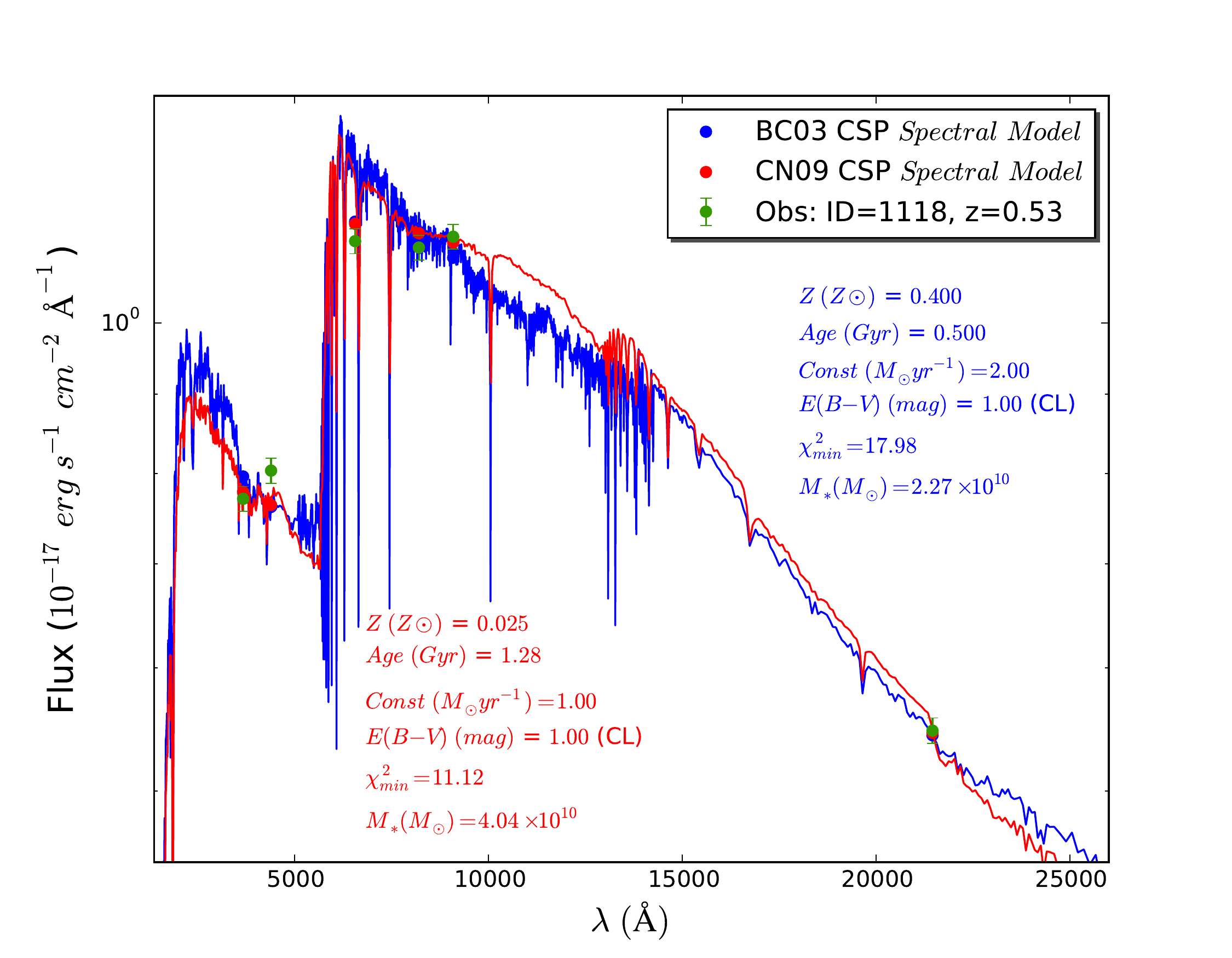} &
    \includegraphics[width=.5\textwidth]{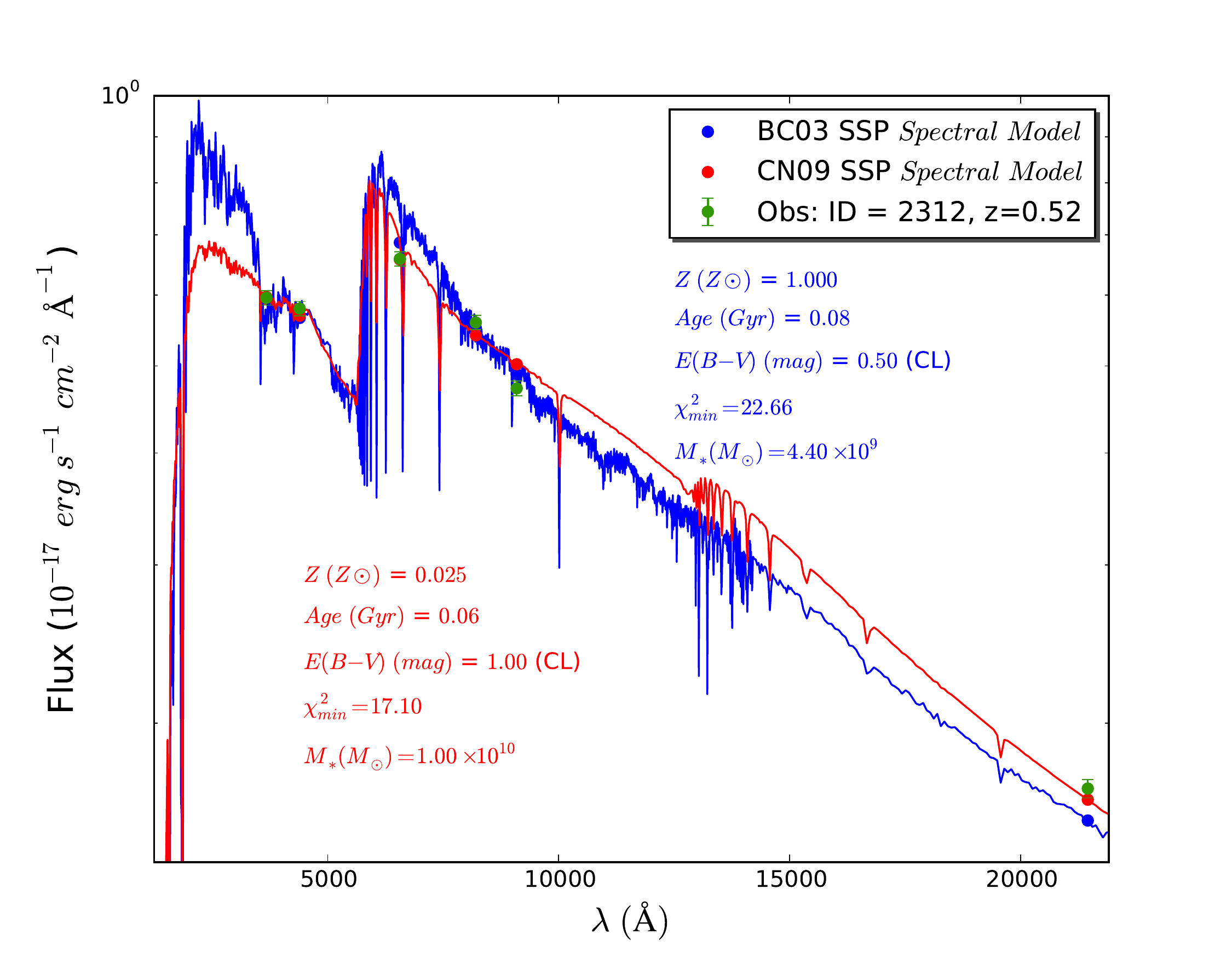}   
   \end{tabular}
   \caption{(continued)}
      \label{fig:A3}
  \end{figure*}
  
  \begin{figure*}
\centering
  \begin{tabular}{@{}cc@{}}
    \includegraphics[width=.5\textwidth]{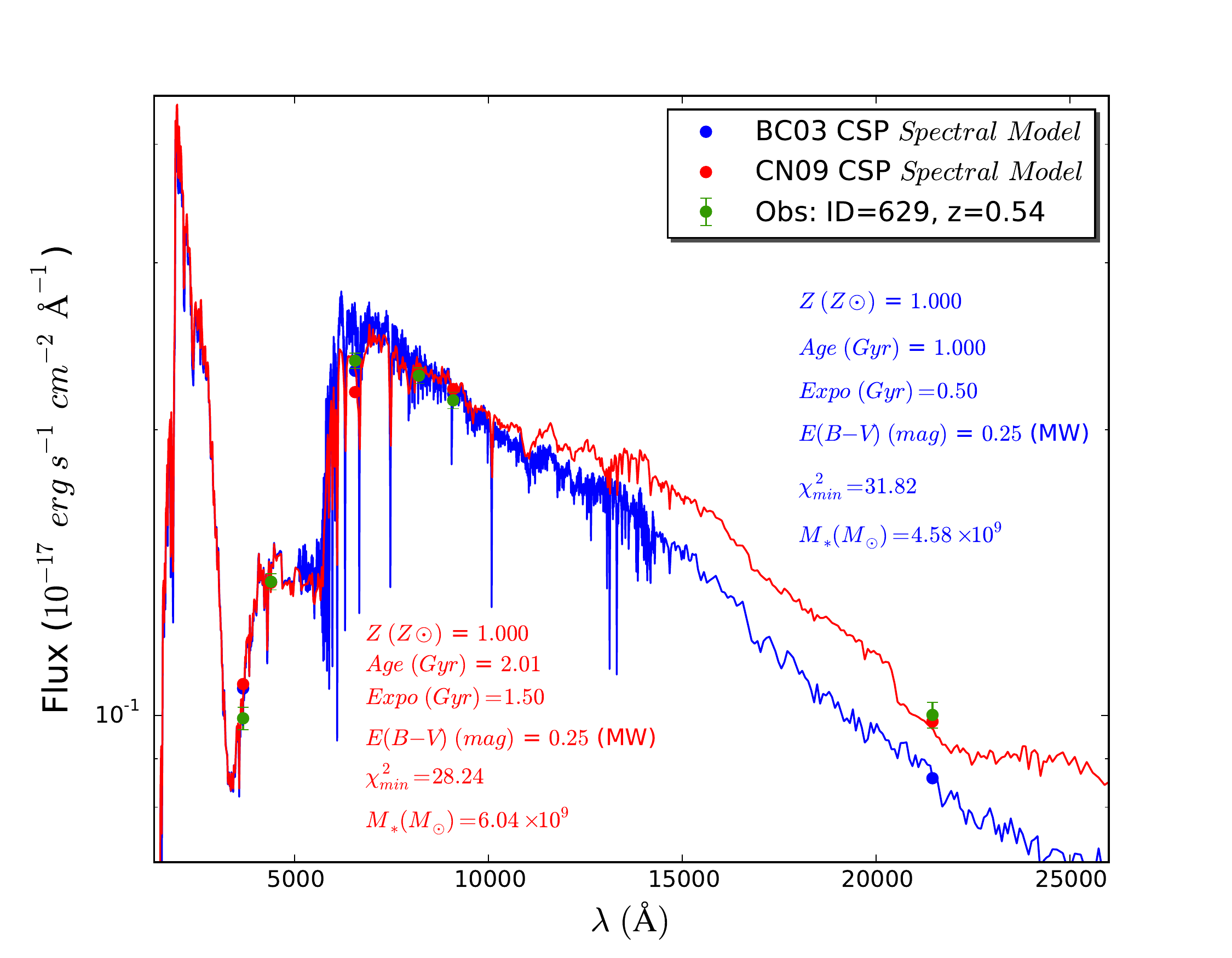} &
    \includegraphics[width=.5\textwidth]{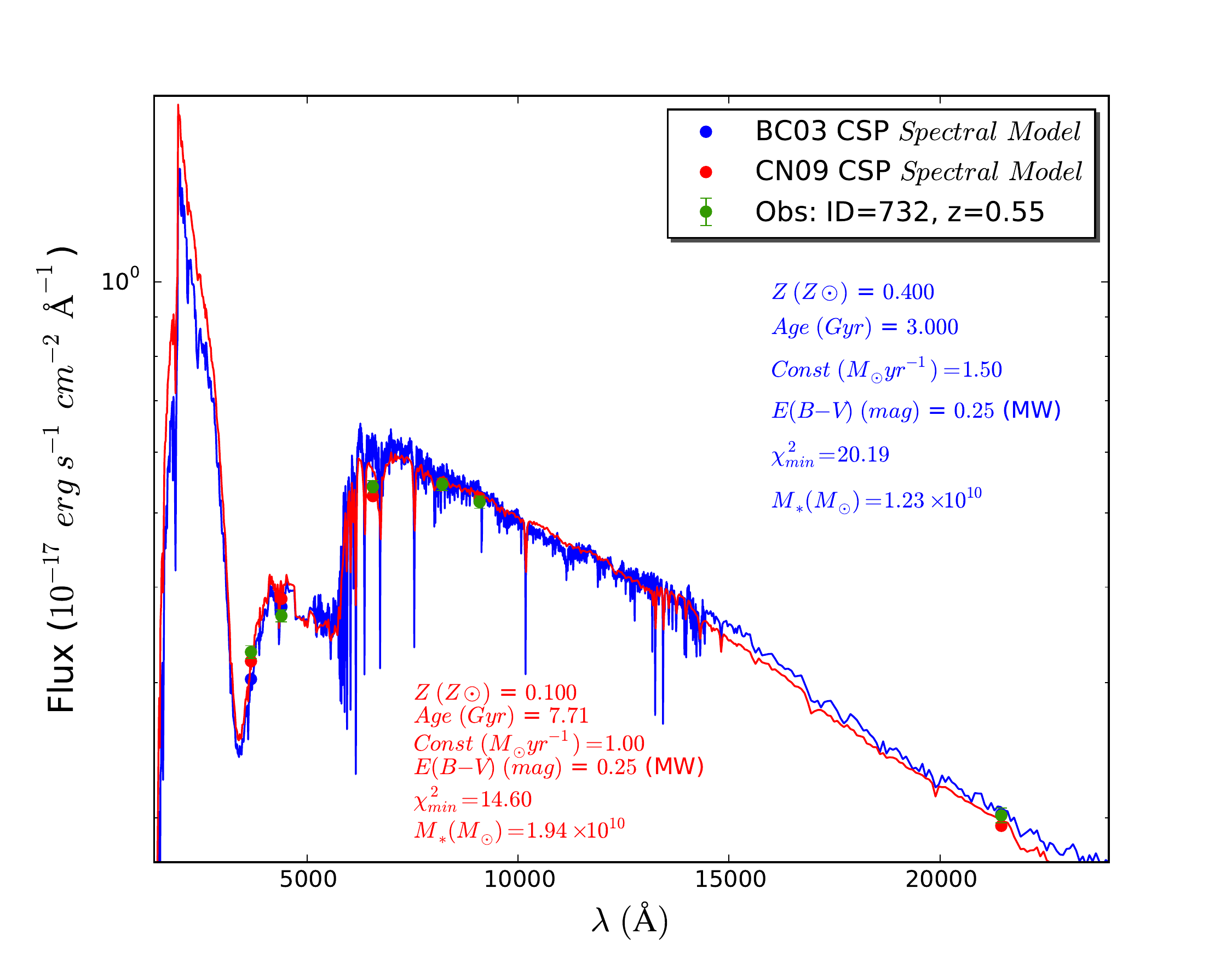}   \\
    \includegraphics[width=.5\textwidth]{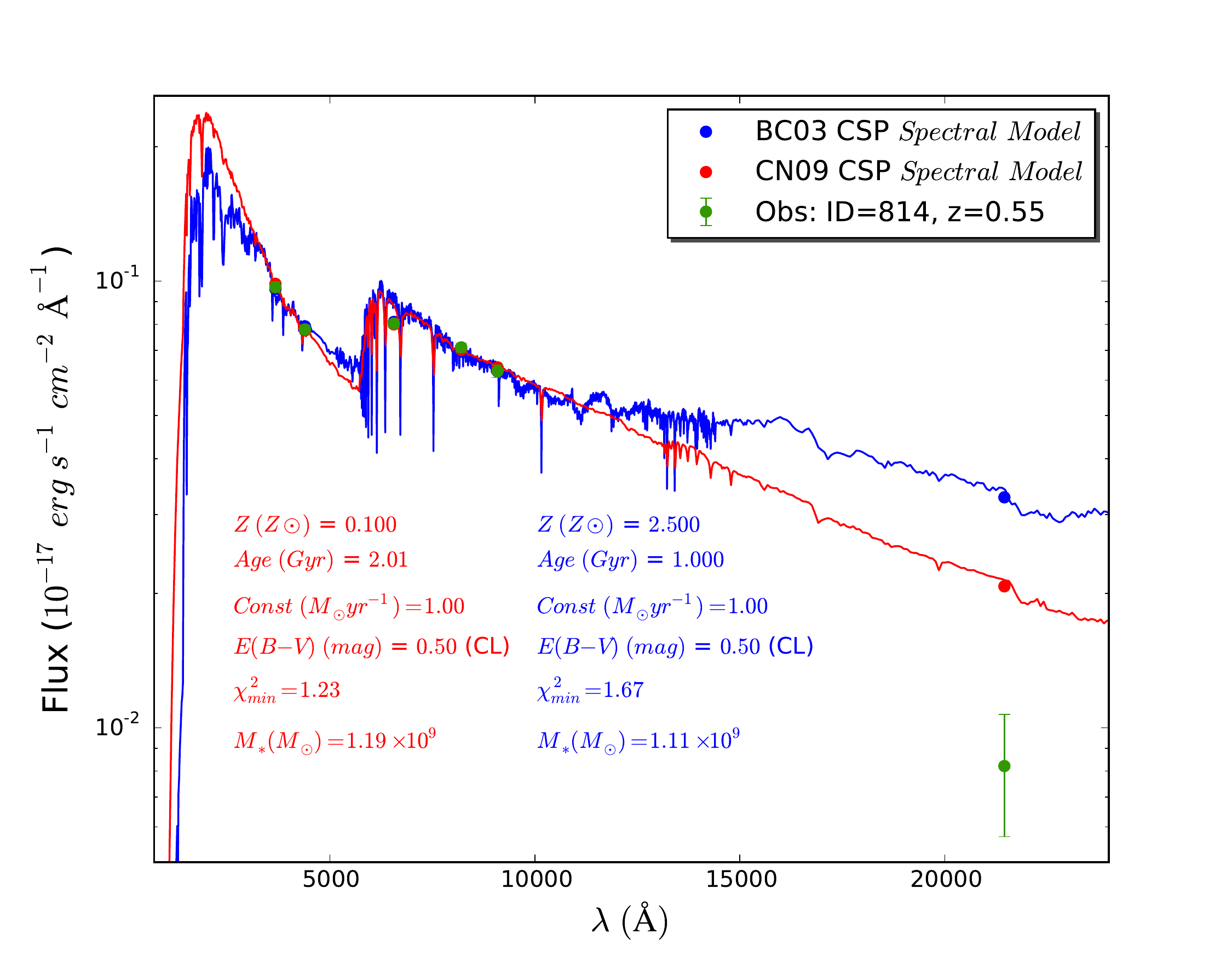}   &
    \includegraphics[width=.5\textwidth]{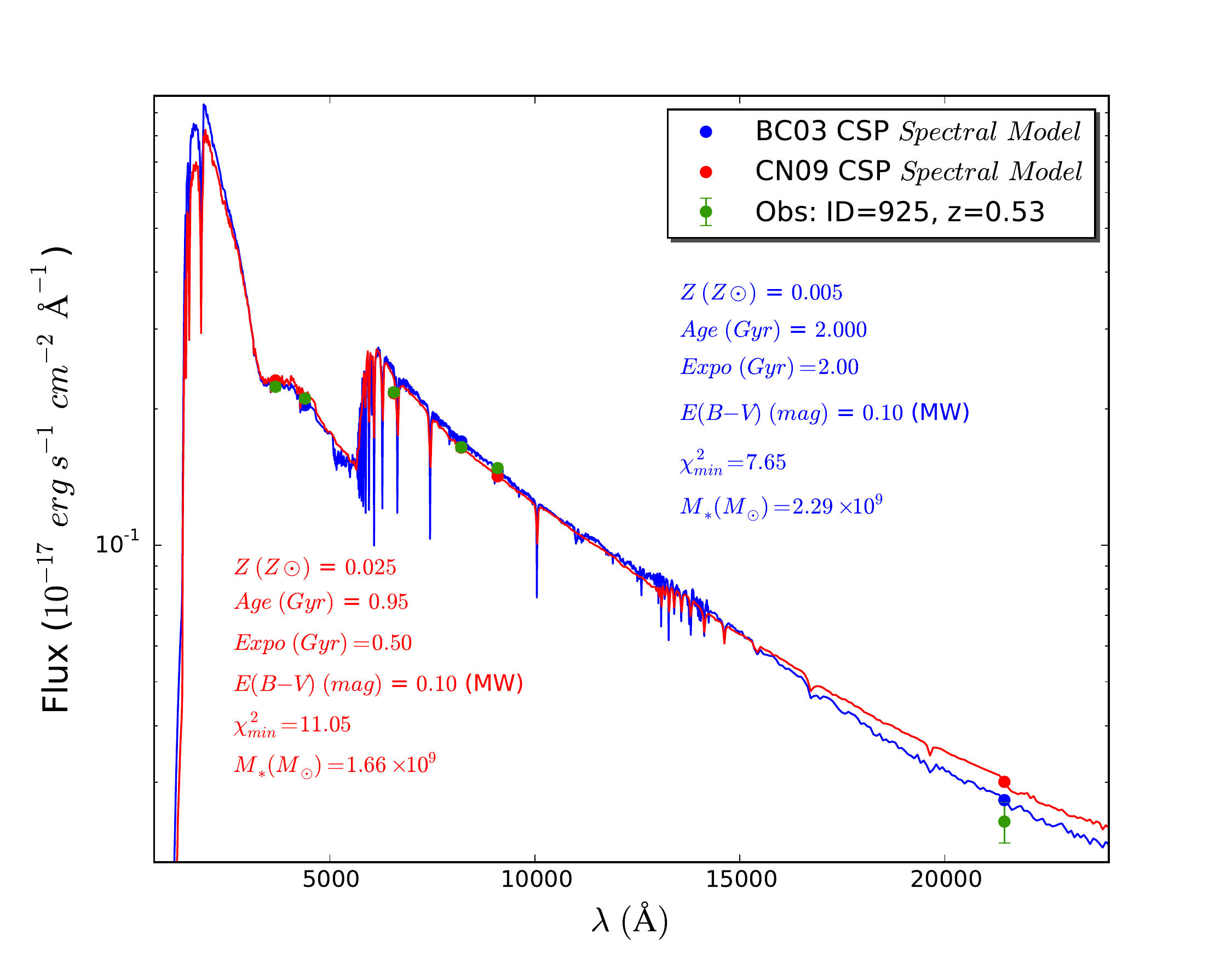}  \\
     \includegraphics[width=.5\textwidth]{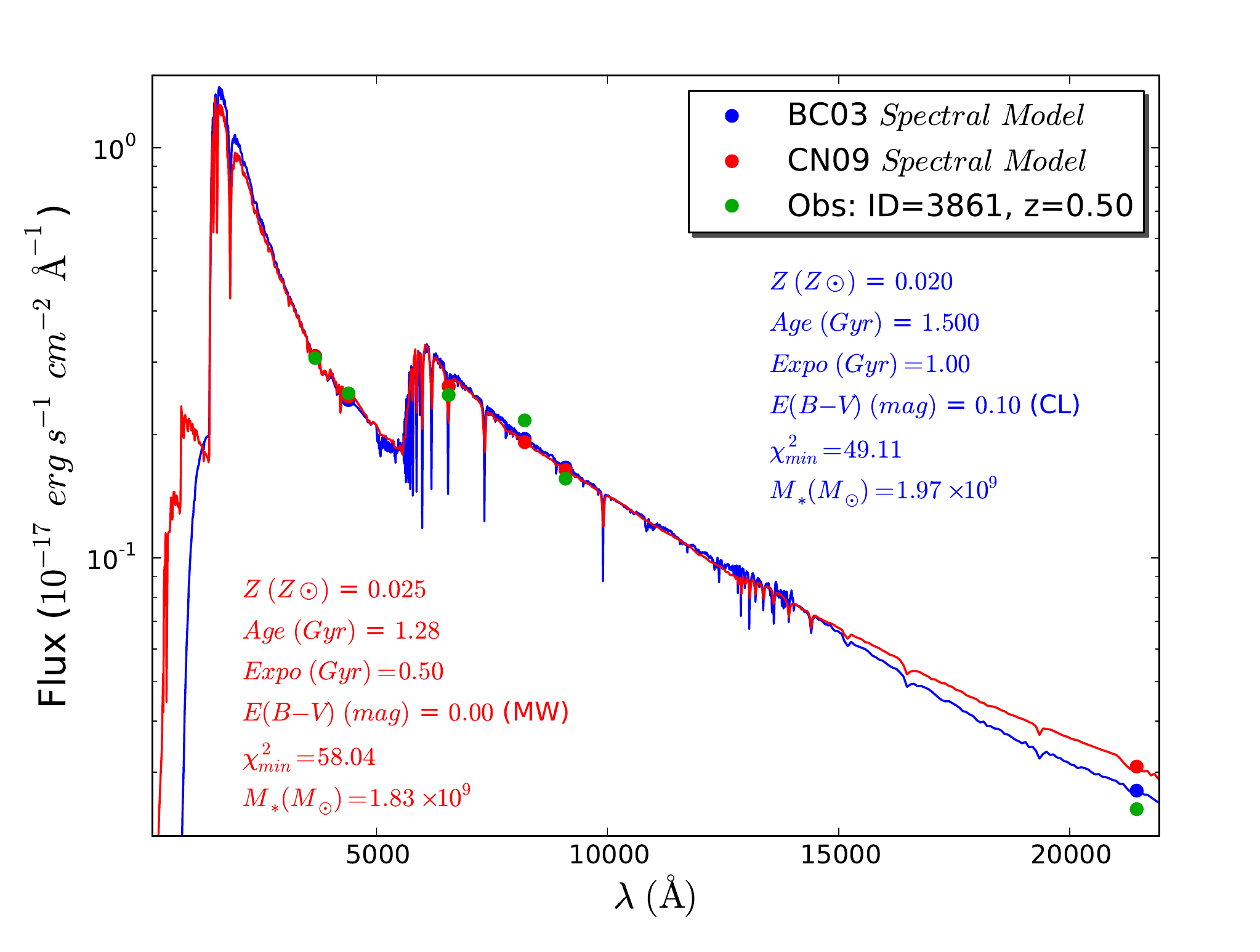} 
   \end{tabular}
   \caption{(continued)}
      \label{fig:A4}
  \end{figure*}


\bsp	
\label{lastpage}
\end{document}